\PassOptionsToPackage{hypertexnames=false}{hyperref}
\documentclass[twocolumn, tighten,twocolappendix]{aastex7}
\graphicspath{figures/}
\usepackage{needspace}
\usepackage{newtxtext,newtxmath}
\usepackage{orcidlink}
\usepackage{booktabs,bigdelim,enumitem}
\setlist{noitemsep} % or \setlist{noitemsep} to leave space around whole list
\usepackage{placeins}
\usepackage{fancyhdr}
\newcommand{\Illustris}{\textsc{Illustris}}
\newcommand{\TNG}{\textsc{TNG}}
\newcommand{\TNGfifty}{\textsc{TNG50}}
\newcommand{\TNGfield}{\texttt}
\newcommand{\AREPO}{\textsc{Arepo}}
\newcommand{\Subfind}{\textsc{Subfind}}
\newcommand{\Sublink}{\textsc{Sublink}}

\newcommand{\Msun}{\ensuremath{\mathrm{M}_\odot}}
\newcommand{\Mstar}{\ensuremath{M_\star}}
\newcommand{\Mdyn}{\ensuremath{M_\mathrm{dyn}}}
\newcommand{\Rdyn}{\ensuremath{R_\mathrm{dyn}}}
\newcommand{\tin}{\ensuremath{t_\mathrm{infall}}}

\newcommand{\exs}{\text{ex situ}}

\newcommand{\ins}{\text{in situ}}

\begin{document}
\title{\vspace{-\baselineskip}\Large Cosmological Simulations of Stellar Halos with Gaia Sausage--Enceladus Analogs:\\Two Sausages, One Bun?}
\shorttitle{Stellar Halos with GSEs}

\author{\rm Dylan Folsom~\orcidlink{0000-0002-1544-1381}}
% \author[0000-0002-1544-1381]{Dylan Folsom}
\email{dfolsom@princeton.edu}
\affiliation{\rm Department of Physics, Princeton University, Princeton, NJ 08544, USA; \href{mailto:dfolsom@princeton.edu}{dfolsom@princeton.edu}}

\author{\rm Mariangela Lisanti~\orcidlink{0000-0002-8495-8659}}
% \author[0000-0002-8495-8659]{Mariangela Lisanti}
\email{dfolsom@princeton.edu}
\affiliation{\rm Department of Physics, Princeton University, Princeton, NJ 08544, USA; \href{mailto:dfolsom@princeton.edu}{dfolsom@princeton.edu}}
\affiliation{\rm Center for Computational Astrophysics, Flatiron Institute, New York, NY 10010, USA}

\author{\rm Lina Necib~\orcidlink{0000-0003-2806-1414}}
% \author[0000-0003-2806-1414]{Lina Necib}
\email{dfolsom@princeton.edu}
\affiliation{\rm Physics Department and Kavli Institute for Astrophysics and Space Research, Massachusetts Institute of Technology, Cambridge, MA 02139, USA}
\affiliation{\rm The NSF AI Institute for Artificial Intelligence and Fundamental Interactions, Massachusetts Institute of Technology, Cambridge, MA 02139, USA}

\author{\rm Danny Horta~\orcidlink{0000-0003-1856-2151}}
% \author[0000-0003-1856-2151]{Danny Horta}
\email{dfolsom@princeton.edu}
\affiliation{\rm Center for Computational Astrophysics, Flatiron Institute, New York, NY 10010, USA}

\author{\rm Mark Vogelsberger~\orcidlink{0000-0001-8593-7692}}
% \author[0000-0001-8593-7692]{Mark Vogelsberger}
\email{dfolsom@princeton.edu}
\affiliation{\rm Physics Department and Kavli Institute for Astrophysics and Space Research, Massachusetts Institute of Technology, Cambridge, MA 02139, USA}
\affiliation{\rm The NSF AI Institute for Artificial Intelligence and Fundamental Interactions, Massachusetts Institute of Technology, Cambridge, MA 02139, USA}

\author{\rm Lars Hernquist~\orcidlink{0000-0001-6950-1629}}
% \author[0000-0001-6950-1629]{Lars Hernquist}
\email{dfolsom@princeton.edu}
\affiliation{\rm Center for Astrophysics, Harvard \& Smithsonian, 60 Garden Street, Cambridge, MA 02138, USA\\\it Received 2024 November 10; revised 2025 February 17; accepted 2025 March 5; published 2025 April 14}

\journalinfo{The Astrophysical Journal, 983:119 (18pp), 2025 April 20}

\shortauthors{Folsom et al.}

\begin{abstract}
\noindent Observations of the Milky Way's stellar halo find that it is predominantly comprised of a radially biased population of stars, dubbed the Gaia Sausage--Enceladus, or GSE. These stars are thought to be debris from dwarf galaxy accretion early in the Milky Way's history. Though typically considered to be from a single merger, it is possible that the GSE debris has multiple sources. To investigate this possibility, we use the \TNGfifty{} simulation to identify stellar accretion histories in 98 Milky Way analogs---the largest sample for which such an identification has been performed---and find GSE-like debris in 32, with two-merger GSEs accounting for a third of these cases. Distinguishing single-merger GSEs from two-merger GSEs is difficult in common kinematic spaces, but differences are more evident through chemical abundances and star formation histories. This is because single-merger GSEs are typically accreted more recently than the galaxies in two-merger GSEs: the median infall times (with 16th and 84th percentiles) are $5.9_{-2.0}^{+3.3}$ and $10.7_{-3.7}^{+1.2}$~Gyr ago for single- and two-merger scenarios, respectively. The systematic shifts in abundances and ages that occur as a result suggest that efforts in modeling these aspects of the stellar halo prove ever-important in understanding its assembly.

\vspace{0.5\baselineskip}\noindent\textit{Unified Astronomy Thesaurus concepts:} \href{http://astrothesaurus.org/uat/2178}{Galactic archaeology (2178)}; \href{http://astrothesaurus.org/uat/1608}{Stellar kinematics (1608)}; \href{http://astrothesaurus.org/uat/598}{Galaxy stellar halos (598)}
\end{abstract}

\section{Introduction}
\label{sec:1}
\nopagebreak
One key prediction of the Lambda cold dark matter~($\Lambda$CDM) cosmological paradigm is that of hierarchical structure formation~\citep{White78,White91}. In this model, the early Universe has a stochastically varying density distribution, in which overdensities serve as the seeds of galaxy formation. These overdensities interact gravitationally, growing by merging and accreting nearby matter. The assembly histories of galaxies like the Milky Way~(MW) depend on the stochastic initial conditions of this density field, and each galaxy will form differently, even if the resulting halos are qualitatively similar. During the assembly process, infalling satellite galaxies are tidally disrupted by the strong gravitational forces of their host, leaving behind debris in ``stellar halos''~\citep{Johnston96,Johnston98,Bullock05}. The debris from these mergers can retain distinctive structure in chemodynamical space, thereby providing a useful approach to both identify and characterize these ancient events with present-day observations~\citep{Robertson05,Font06,Font11}.

\thispagestyle{fancy}

Such observations are well underway, thanks to the now-completed RAVE~\citep{Steinmetz20}, SEGUE~\citep{Rockosi22}, and APOGEE~\citep{Abdurrouf22} surveys; the ongoing efforts from LAMOST~\citep{Cui12}, H3~\citep{Conroy19}, GALAH~\citep{Buder21}, \emph{Gaia}~\citep{GaiaCollaboration22}, and DESI~\citep{Cooper23}; and planned surveys such as 4MOST~\citep{Helmi19}, WEAVE~\citep{Jin24}, and the LSST~\citep{Ivezic19}. The result of these efforts thus far is an extensive picture of the inner~($\lesssim 50$~kpc) stellar halo of the MW. 

The stellar halo is rich in structure, as summarized in~\cite{Helmi20}; it appears to be almost entirely comprised of debris from merging satellites~\citep{Deason15,Deason18,Haywood18,Kruijssen19,Naidu20,Malhan22,Malhan24} and the \ins{} stars they heat out of the disk \citep[][though see also \citet{Amarante20}]{Zolotov09,Helmi18,Belokurov20}. The most significant accreted contribution to the inner halo is dubbed the \textit{Gaia} Sausage--Enceladus, or GSE~\citep{Belokurov18,Helmi18}---see \citet{Deason24} for a recent review. The GSE stellar debris comprises more than 50\% of the \exs{} stars within 30~kpc of the Galactic center and has significantly radially-biased velocity anisotropy \citep{Deason18,Myeong18,Lancaster19,Necib19a,Iorio21,Naidu21}. 

Despite its well-studied chemodynamical properties, the underlying nature of the GSE debris is still being brought to light: many studies \citep{Bonaca20,Feuillet21,Grunblatt21,Hasselquist21,Horta24} find stars which are as young as $\sim 6$, rather than $\gtrsim 10$,~Gyr old, but otherwise appear consistent with the GSE stellar population. One potential explanation for this is slow orbital decay allowing for ongoing star formation in the progenitor~\citep{Bonaca20}; however, another hypothesis is an additional dwarf galaxy component to the GSE debris. Rather than being from a single merger, these stars may have been contributed by multiple independent mergers. This hypothesis has been explored by~\citet{Donlon22}, who argue that the structures present in chemodynamical spaces are more easily explained by multiple mergers than the single-merger GSE paradigm, a claim also consistent with a later infall time for GSE debris \citep{Donlon20,Donlon24}. 

Regardless of its origin, the ubiquity of the GSE debris in the inner halo makes it a useful testbed for assessing the consistency between expectations from hierarchical structure formation in the $\Lambda$CDM paradigm and observations of our Galaxy's structure. In particular, models of galaxy formation must be consistent with GSE-like structures, in which a galaxy's stellar halo is overwhelmingly composed of a population of radially anisotropic (RA) stars, whether this population is from a single large merger or from multiple smaller mergers. 

Assessing the prevalence of GSE-like debris in a cosmological context and discerning its origin requires an understanding of the range of possibilities allowed for the accretion history of MW-like galaxies in a CDM universe, which can be substantial due to cosmological variance. Fortunately, state-of-the-art large-volume, high-resolution simulations are expanding the possibility of quantifying the spread of expectations by providing substantial samples of MW-like galaxies that evolved in a cosmological setting. In this work, we use the \Illustris\TNG{} project~\citep{Nelson21}, and the high-resolution \TNGfifty{} simulation \citep{Nelson19,Pillepich19} in particular.\footnote{Other comparable simulations include NewHorizon~\citep{Dubois14} and FIREbox~\citep{Feldmann23}. See also \citet{Vogelsberger20} for a recent review of cosmological simulations.} Encompassing a volume of (50~comoving~Mpc)$^3$, \TNGfifty{} includes $\mathcal{O}(100)$ halos near the MW's virial mass, resolving their subhalos down to nearly $10^7~\Msun$. This simulated sample of MW-like galaxies provides an unprecedented opportunity to characterize the spread in stellar properties in galaxies like our own, leveraging the large volume of the simulation and correspondingly large sample size. Within this context, it is possible to better understand how common or rare the MW is among the range of possibilities. Moreover, it allows one to study large samples of GSE-like events, understanding how likely they are to form and from how many mergers.

\addtolength{\topmargin}{1cm}
\addtolength{\headsep}{-1cm}
\lhead{\footnotesize \textsc{The Astrophysical Journal}, 983:119 (18pp), 2025 April 20}
\rhead{\footnotesize Folsom et al.}

Previous simulations have focused either on reproducing the properties of the GSE merger directly \citep[e.g.][]{Koppelman20,Naidu21,Amarante22,Rey23,Buch24} or on selecting GSE-like events from a larger volume \citep[e.g.][]{Bignone19,Fattahi19,Mackereth19,Elias20,Evans20,Grand20,Dillamore22,Belokurov23,Garcia-Bethencourt23,Orkney23,Rey23}. The majority of these works look for GSE analogs in $\mathcal{O}(10)$ MW analogs at most, limited in statistics by the scope of the simulation. Other works have studied GSE analogs with sample sizes comparable to (or greater than) our own, namely, in \textsc{artemis}~\citep{Dillamore22}, \textsc{coco}/\textsc{colour}~\citep{Bose20}, EAGLE~\citep{Bignone19,Evans20}, and \Illustris{}~\citep{Elias20}. Each of these works uses a sample of MW analogs comparable (or greater) in size to our own, but the present work provides a combination of large statistical power, state-of-the-art magnetohydrodynamics, and detailed phase-space criteria for GSE debris not present in previous studies. 

In this work, we construct the assembly history of each MW-like galaxy in TNG50, identifying the mergers that contribute RA debris to its inner halo, as expected of the GSE. Roughly 30\% of the TNG MWs host GSE-like events, although a smaller fraction (13\% of MWs with GSE-like events, or 4\% overall) match the extreme velocity anisotropy observed in our Galaxy's GSE. We demonstrate that there is a reasonable probability that GSE-like debris is built from more than a single merger within a $\Lambda$CDM context and explore whether these merger components can be distinguished with kinematic, abundance, or age information.

This paper is organized as follows: \autoref{sec:2} introduces \TNG{}, the criteria used to select the sample of MW analogs, the procedure used to recover the merger history for each galaxy in the sample, and the tagging of individual star particles as belonging to a particular merger. \autoref{sec:3} discusses the stellar halos across the sample and establishes a subsample of halos with RA stellar debris. \autoref{sec:4} discusses the kinematic properties of this debris, and \autoref{sec:5} discusses their star formation histories (SFHs) and chemical properties. Furthermore, \autoref{sec:6} summarizes the primary conclusions of this work. \autoref{app:A} gives the IDs of MW analogs within the simulation, \autoref{app:B} provides additional technical details regarding the methodology~(\autoref{sec:2}), and \autoref{app:C} provides additional figures for each MW analog in the sample. 

\Needspace*{4\baselineskip}
\section{Methodology}
\label{sec:2}
\nopagebreak

\subsection{Simulations}
\label{sec:2.1}
\nopagebreak
The \Illustris\TNG{} project~\citep{Nelson19} is a suite of magnetohydrodynamic cosmological simulations performed with the moving-mesh code \AREPO{}~\citep{Springel10} and is a successor to \Illustris{}~\citep{Vogelsberger14,Vogelsberger14a}.\footnote{All \Illustris\TNG{} data, including post-processed catalogs, are available online at \href{https://tng-project.org/}{https://tng-project.org/}.} This work utilizes the highest-resolution simulation, \TNGfifty{} \citep[][hereafter simply \TNG{}]{Nelson19a,Pillepich19}, which covers a volume of $(51.7~\text{Mpc})^3$ from a redshift of $z=127$ to the present day according to cosmological parameters from the \citet{PlanckCollaboration16}, viz. $\Omega_{\Lambda} = 0.6911$, $\Omega_m =0.3089$, $\Omega_b=0.0486$, $\sigma_8=0.8159$, $n_s=0.9667$, and $h=0.6774$. The simulation uses a dark matter particle mass resolution of $4.5\times10^5~\Msun$ and a baryonic mass resolution of $8.5\times 10^4~\Msun$.

Certain time steps of the simulation are saved as ``snapshots,'' and within these snapshots, a halo-finding algorithm is used to track the growth and evolution of structure. This process is broken down into two steps. First, structures are identified with a friends-of-friends~(FoF) algorithm \citep{Davis85}, which is run only on the dark matter particles; other particle types (gas, stars, and black holes) are associated with the same FoF group as their nearest dark matter particle. These groups generally consist of a large central host halo and the subhalos in its potential well, though the FoF algorithm does not identify subhalos, nor does it require that the group be gravitationally bound. Next, bound structures (i.e., dark matter halos) are found within FoF groups using the \Subfind{} algorithm~\citep{Springel01,Dolag09}, which considers all particle types.

There are a number of masses defined in the \TNG{} FoF and \Subfind{} catalogs relevant to this study. Each FoF group has a mass denoted $M_{200}$, which is the total mass within a sphere of radius $R_{200}$ drawn around the most bound particle of the FoF group such that the average density of that sphere is $200$ times the critical density of the Universe. Since this is computed for FoF groups only, halos that are not the central objects in their group (i.e., satellites or other spatially adjacent halos linked to the central halo by the FoF algorithm) do not have this quantity computed. However, each \Subfind{} halo does have a total bound mass, hence denoted \Mdyn{}.\footnote{We define a radius \Rdyn{} from \Mdyn{} such that a sphere with radius \Rdyn{} and density 200 times the critical density has mass \Mdyn{}, the same relation between $R_{200}$ and $M_{200}$.} This is the total mass of all particles bound to the halo, excluding the mass bound to any satellite halos. The contribution to \Mdyn{} that comes from star particles is denoted \Mstar{}. Since \Mdyn{} and \Mstar{} are defined for all halos, not only those that are central in the FoF groups, these are the masses considered in this work. 

Because halos are identified in each individual snapshot, it is nontrivial to track the same physical object from snapshot to snapshot. To this end, the \Illustris\TNG{} project provides a ``merger tree,'' which connects persistent physical structure between snapshots, recording the IDs given to the same structure at different points in time. Earlier snapshots hold ``progenitors'' of the later snapshots' ``descendant'' halos. In \TNG{}, these trees are found by the \Sublink{} algorithm \citep{Rodriguez-Gomez15}, though in this study, we modify the \Sublink{} trees in cases where the algorithm fails to properly identify a descendant halo. More information about \Sublink{} and our modifications is given in \autoref{app:B}.

\subsection{Selection Criteria for MW Analogs}
\label{sec:2.2}
\nopagebreak

\begin{table}
  \renewcommand{\arraystretch}{0.8}
  \setlength{\baselineskip}{4pt}\selectfont
  \caption{Selection Criteria for Milky Way Analogs}
  \begin{tabular}{l@{\hspace{-2em}}c}
  \toprule
  Selection Cut & Number of Halos \\
  \midrule
  $\Mstar{} \in [4,\,7.3] \times10^{10}~\Msun$ & 131\\
  $>500$~kpc from halo with larger $\Mdyn$ & 109\\
  $>1$~Mpc from halo with $\Mdyn{} > 10^{13}~\Msun$ & 98\\
  \midrule
  Single RA merger & 21\\
  RA pair & 11\\
  \bottomrule
  \end{tabular}
  \tablecomments{ The top portion of the table enumerates the selection of MW analogs within the \TNG{} volume (see~\autoref{sec:2.2}). The cuts listed in the table are applied sequentially, with each diminishing the sample. Below this, the number of halos with RA debris is shown, split into subsets in which the debris is due to one or two mergers (see~\autoref{sec:3.2}). Five of the 11 analogs with an RA pair have two choices of merger pairs that satisfy the selection requirements---see discussion in the main text.}
  \label{tab:1}
\end{table}

A series of selection criteria are required to isolate galaxies within TNG that most closely resemble the MW. These analogs are chosen from the $z = 0$ snapshot based on their stellar mass and two additional isolation criteria.
In particular, they must
\begin{enumerate}
  \item have \Mstar{} within $4\times 10^{10}~\Msun$ and $7.3\times 10^{10}~\Msun$,
  \item be farther than 500~kpc from any halo with an \Mdyn{} larger than that of the candidate halo, and
  \item be farther than 1~Mpc from any halo with $\Mdyn > 10^{13}~\Msun$. 
\end{enumerate} 
The result of these selections is summarized in \autoref{tab:1}. Note that there are no particular requirements on the MW itself other than its stellar mass; we do not select for any aspects of the Galaxy's stellar morphology, even those that may be influenced by the assembly history. Further, there need not be a Large Magellanic Cloud (LMC) analog, an analog of M31, or other Local Group features. Such a local environment may be correlated with its formation history, and being agnostic to this may introduce systematic biases in the sample.

The criterion on stellar mass is based on 68\% confidence intervals quoted in the literature~\citep[with \citet{Bland-Hawthorn16} and \citet{Licquia15} providing the widest upper and lower bound, respectively]{Flynn06,McMillan11,Licquia15,Bland-Hawthorn16,Licquia16,McMillan17,Cautun20}. This ensures that the selected galaxies have a stellar mass comparable to the MW, though the range is intentionally left broad. There are a total of 131 galaxies in the \TNG{} volume that satisfy this requirement.

The first of the two isolation requirements ensures that the MW analog is not too close to any larger galaxy, with ``large'' determined by the total bound mass \Mdyn{}. In the Local Group, the nearest larger galaxy is M31, with a mass around $(1$--$2)\times 10^{12}~\Msun$~\citep{Patel17,Kafle18,Benisty22,Sawala23,Villanueva-Domingo23} and distance $\sim760$~kpc from the MW~\citep{Li21,Lee23}. The distance to an M31 analog in \TNG{} is somewhat arbitrary, however. This study considers the final snapshot as $z = 0$, but this need not be the case. Snapshots in \TNG{} at late times are spaced approximately 170~Myr apart. Given that M31 is approaching at a velocity of $\sim 110~\mathrm{kpc}\,\mathrm{Gyr}^{-1}$~\citep{VanDerMarel12}, the distance to M31 could vary by tens of kiloparsecs due to the choice of snapshot. Therefore, the distance to a nearby partner is relaxed; in keeping with \citet{Pillepich23}, 500~kpc is taken as the minimum separation between the MW and the nearest larger galaxy. 109 (83\%) of the galaxies that pass the initial stellar mass threshold also pass this isolation requirement. 

The second of the two isolation requirements is on a larger scale: the Local Group is known to be somewhat isolated, with no large galaxy groups in the nearby Universe~\citep{Karachentsev05}. To this end, the MW analogs are required to be more than 1~Mpc from any halo with a total bound mass typical of massive groups, $\Mdyn \sim 10^{13}~\Msun$. A total of 98 (90\%) of the 109 galaxies that satisfy the first isolation requirement also satisfy this condition, and these 98 halos comprise the final sample of MW analogs. Their \TNG{} \Subfind~ID numbers are provided in Appendix~\ref{app:A}. 

It is important to note that the \TNG{} box contains two Virgo-mass clusters ($\Mdyn \gtrsim 10^{14}~\Msun$) within the (51.7~Mpc)$^3$ volume. The Virgo cluster itself is $\sim16$~Mpc from the MW \citep{Cantiello24}, but few of the MW analogs are this isolated: 79 (81\%) of the 98 halos fall within 16~Mpc of one of the two largest clusters, and three analogs are close enough to belong to the FoF group of the largest cluster. Of the MW analogs that are found to host GSE-like debris (see~\autoref{sec:3.2}), the closest is 3.6~Mpc from one of the large clusters.

The criteria used in this work are similar to, though distinct from, those chosen by \citet{Pillepich23}, who established a catalog of MW and M31 analogs in the \TNG{} volume. Though the majority (85\%) of the sample defined above is included in their catalog, there are three primary differences in selection criteria.

First, this work focuses on MW analogs in particular, ignoring M31-like galaxies. Thus, the stellar mass criterion used here targets MW-mass galaxies alone. Furthermore, the stellar mass criterion pertains to the total stellar mass, not only the mass within 30~kpc; \citet{Pillepich23} use the mass within 30~kpc to facilitate comparison to M31. All of the galaxies in our sample satisfy the \citet{Pillepich23} stellar mass criterion, but the restricted range of \Mstar{} used in this work, which excludes M31, contributes in large part to the smaller sample size. 

Second, the Mpc-scale isolation criterion is performed on the level of \Subfind{} halos rather than FoF groups in a way that is (for this sample) strictly looser. \citet{Pillepich23} require that the MW resides in a FoF group whose host has an $M_{200}$ mass less than $10^{13}~\Msun$. However, we allow the MW analog to be in an FoF group with a massive host so long as the analog is sufficiently far~($>1$~Mpc) from the host. This allows for MW analogs that have been associated with a distant FoF group host to be included in the study while still ensuring the large-scale environment is empty of such large halos. Of the halos in our sample, four are excluded from~\citet{Pillepich23} due to their FoF group mass, and a further five are excluded because they are near a galaxy with $M_\star(<30~\mathrm{kpc}) > 10^{10.5}~\Msun{}$. 

Third, this study makes no requirements of the analogs' stellar morphology, whereas \citet{Pillepich23} require a stellar disk that is defined either in terms of the ellipticity of the stellar mass distribution or by visual inspection. The focus of this study is on accreted stellar populations, rather than the \ins{} stars that would comprise the majority of a disk component. While we have not checked this requirement directly, we infer that the MW analogs selected in this work do predominantly satisfy the morphology criterion: only six of our analogs are excluded from \citet{Pillepich23} even though they satisfy the other criteria used in that work. 

\subsection{Recovering the Merger History}
\label{sec:2.3}
\nopagebreak
To study the assembly history of each MW analog's \exs{} stellar population, we build a pipeline that appropriately identifies the progenitor for each star in its halo. The key points of this procedure are summarized in this subsection, with further details provided in \autoref{app:B}.

The stars considered in this study are those that are either~(i)~bound directly to the MW analog at $z = 0$, or (ii)~bound to a noncosmological, primarily gaseous overdensity identified by the clustering algorithm as an independent gravitationally bound structure within the MW analog. This latter case is noted in the \TNG{} catalog under the \TNGfield{SubhaloFlag} field, as described on the data release page.\footnote{\href{https://www.tng-project.org/data/docs/background/\#subhaloflag}{https://tng-project.org/data/docs/background}} We consider all such overdensities within the MW analog's \Rdyn{} and add their stars to the sample. 

For a star in the MW analog, we locate its unique particle ID in every snapshot of the simulation and record the halos to which it is bound.\footnote{We exclude snapshots before the formation of the star particle, as well as snapshots in which \Subfind{} does not find the particle to be gravitationally bound to any halo.} With this information, we associate stars with the mergers that contributed them, accounting for both mergers that are completely disrupted and those that survive to $z = 0$ (see \autoref{app:B} for more information). The masses \Mdyn{} and \Mstar{} are recorded for each merger at every snapshot. Each merger is assigned an infall time, which is the last time the halo crosses its MW analog's \Rdyn{}. 

Stars are tagged as originating from the merger to which they are bound for the greatest number of snapshots, with preference given to the one with the largest peak mass in cases of ambiguity.\footnote{Ambiguous cases are rare---see \autoref{app:B}.} For a star to be considered \exs{}, the following criteria must be met:
\begin{enumerate}
  \item the star must be bound to a halo identified as a merger for at least one snapshot, and
  \item in the first snapshot after the star formed, it must be bound to a halo that is not the MW analog itself, but that is flagged as cosmological in origin according to the \TNGfield{SubhaloFlag}.
\end{enumerate}
These criteria prioritize the purity of the \exs{} stellar sample. Another potential definition of \exs{} material is a simple cut on the distance from the host at the time the star particle formed, but this is more challenging to apply to the TNG hosts whose stellar disks (and therefore \ins{} populations) are often quite extended \citep{Pillepich23}. As a result, typical scales used for birth distance selections may not be easily applicable. The second \exs{} criterion used in this work sidesteps this complication but is in the same spirit, as the \exs{} stars must not be bound to the MW analog at the time of their formation.

\begin{figure*}
  \centering 
  \includegraphics{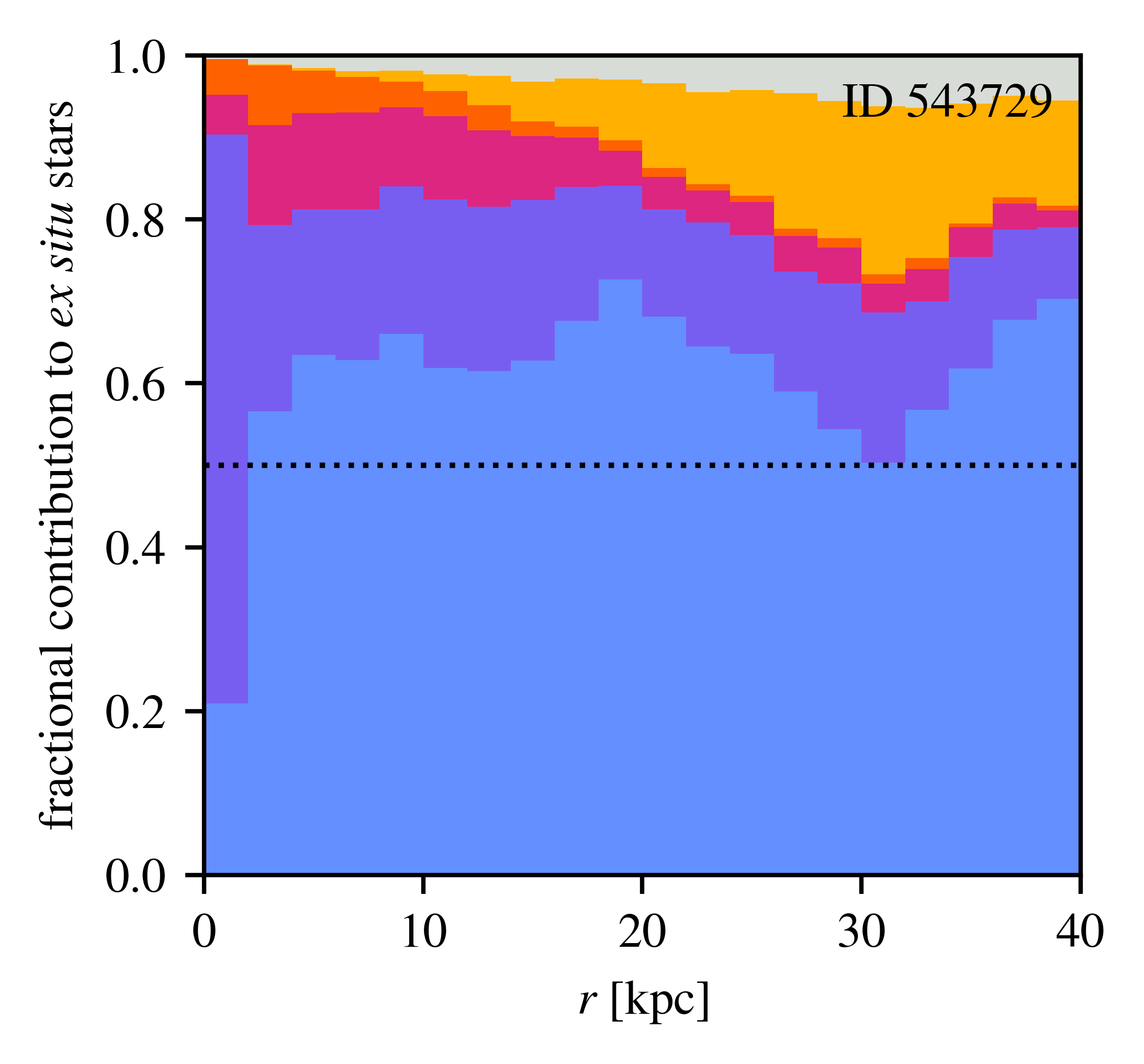}
  \includegraphics{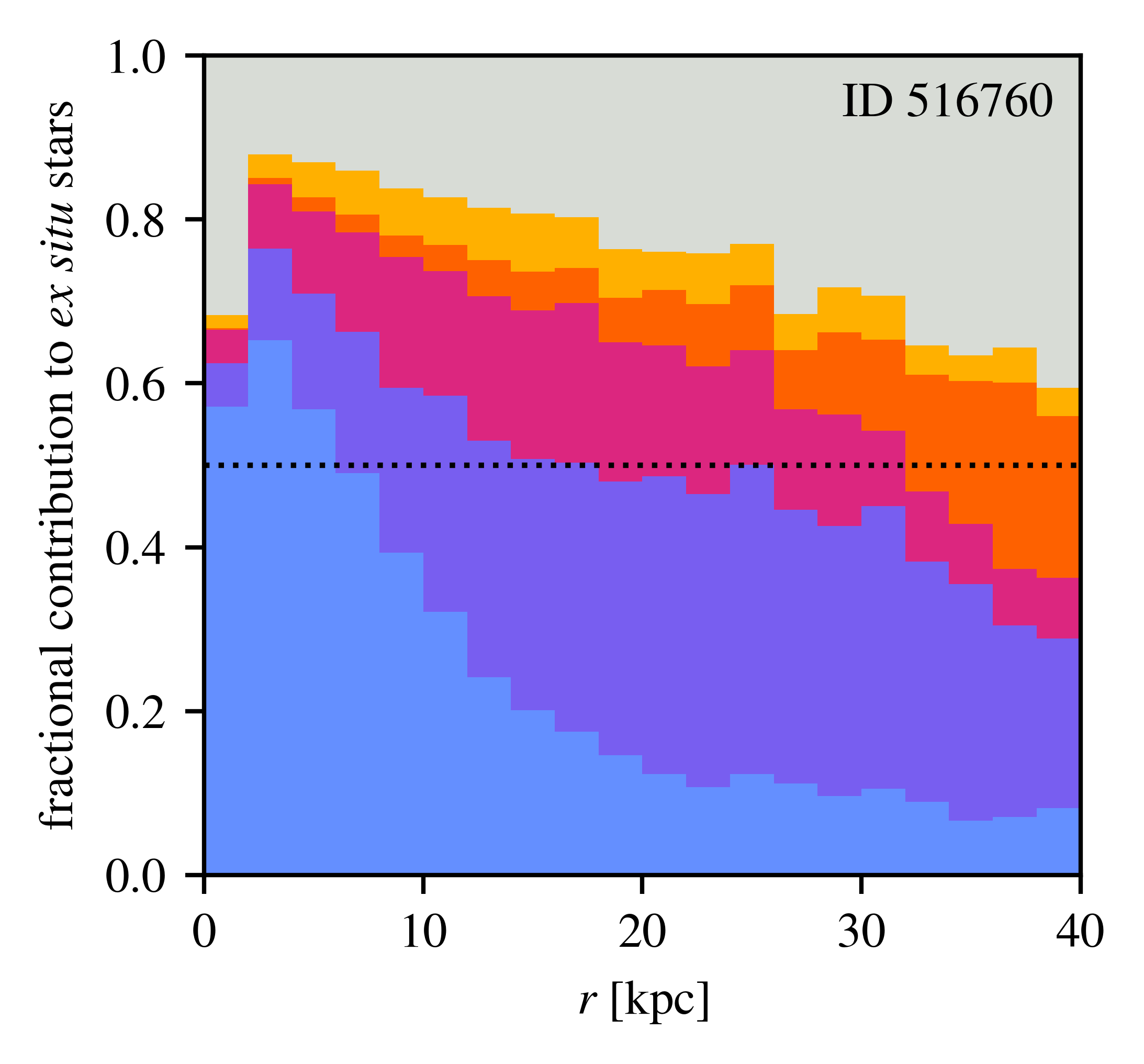}
  \includegraphics{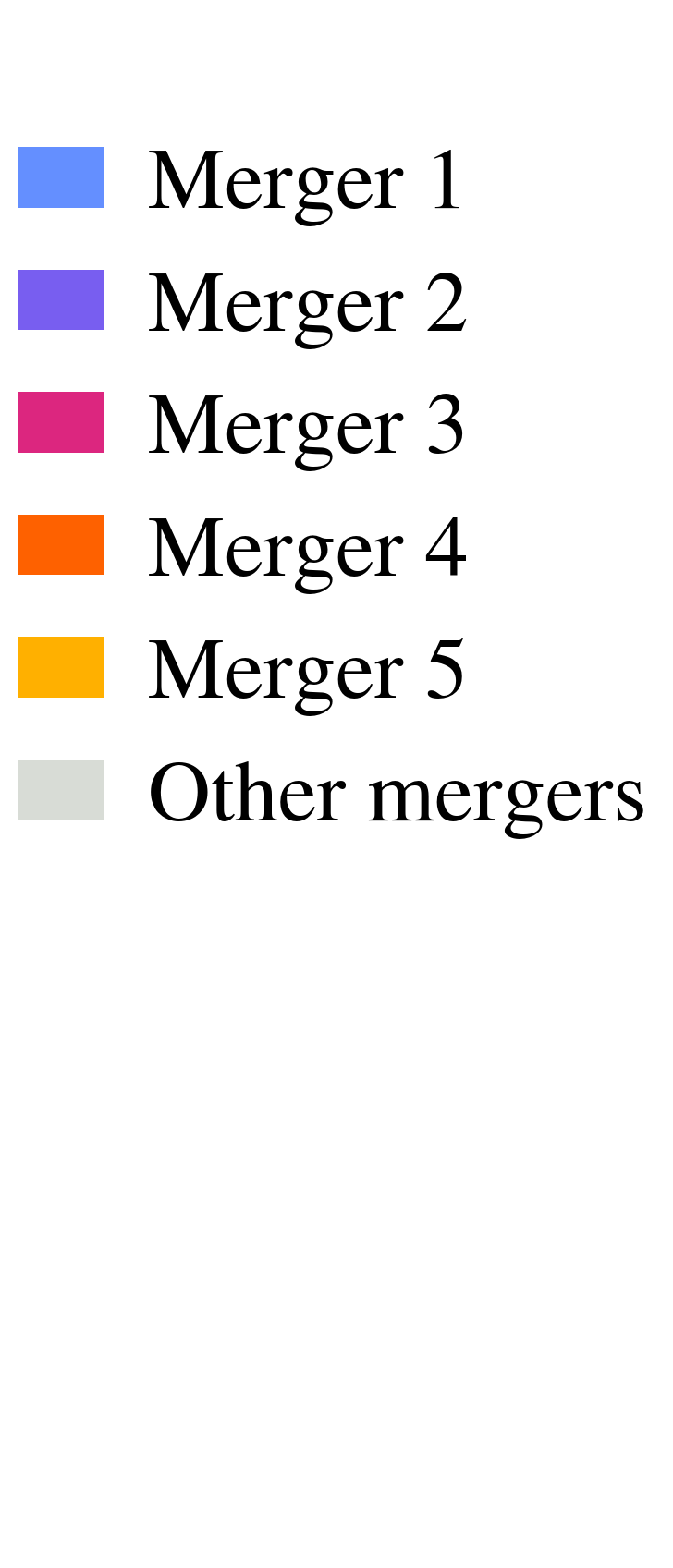}
  \caption{\emph{(Left:)}~The composition of the \exs{} stellar halo for a particular MW analog (ID 543729) as a function of galactocentric distance $r$. The five largest galaxies contributing to the \exs{} stellar population are shown as individual colored bands, sorted by the contribution within 40~kpc. The galaxy that contributes the most \exs{} stellar mass is colored in blue, the second in purple, the third in magenta, the fourth in orange, and the fifth in gold. Stars contributed by all other mergers are collected into the light gray band at the top. This MW analog's \exs{} population is primarily from one large merger with infall $M_\mathrm{dyn} \sim 8\times10^{10}~\Msun$. All other mergers contribute fewer stars to the accreted population. The horizontal line indicates 50\%, the threshold used throughout the work to denote significant mergers. \emph{(Right:)}~The same as the left panel but for ID 516760. This halo hosts a pair of mergers (blue and purple) that, taken together, comprise more than 50\% of the \exs{} stellar material within 40~kpc.}
  \label{fig:1}
\end{figure*}

\Needspace*{4\baselineskip}
\section{Stellar Halos in TNG}
\label{sec:3}
\nopagebreak
A population of 98 MW-like galaxies passes the cuts established in \autoref{sec:2.2}. These galaxies have total masses of\footnote{Logarithms are taken to be base ten throughout this work.} $\log(\Mdyn/\Msun) = 12.05^{+0.17}_{-0.11}$, corresponding to $\Rdyn = 220^{+30}_{-18}$~kpc (16th, 50th, and 84th percentiles). Generally, these galaxies are central to their FoF groups; however, four analogs are part of cluster-scale groups, though they still satisfy the isolation criteria.

All MW analogs are relatively devoid of large satellites, with only 1~(11) analog hosting a satellite of $\Mdyn > 10^{10}~\Msun$ within 50 (100)~kpc.
This is in relative agreement with both theoretical work---e.g.~\citet{Boylan-Kolchin10,Busha11,Bose20,Buch24}---and observations---e.g.~\citet{Liu11,Tollerud11,Mao24}---which find that about 10\% of MW halos host such bright dwarfs (though varying mass and distance criteria shift this number). 

In addition to searching for analogs with LMC partners, one can also classify those that experienced a GSE-like merger(s). In practice, this is more challenging, as it requires tracking the origin of the disrupted material within the \TNG{} halos, rather than simply querying surviving satellites. The merger identification algorithm developed in this work (\autoref{sec:2.3}) enables one to build such histories. This section describes the accreted stellar material in these MW analogs, focusing on identifying those with RA debris expected from GSE-type events. Note that, henceforth, ``stellar halo'' is used to mean ``accreted stellar population,'' excluding \ins{} stars and with no restriction on kinematics.

\subsection{Composition of the Stellar Halos}
\label{sec:3.1}
\nopagebreak
Observations of the MW's stellar halo indicate that the primary source of accreted stellar material in the MW is a small number of dwarf galaxies, with the largest identified component, the GSE, comprising as much as 50\% of the accreted stars within 40~kpc\footnote{This is true as a whole but not necessarily differentially. Specifically, at small radii ($\lesssim 5$~kpc), stars from other accreted structures, such as Heracles~\citep{Horta21}, contribute more than the GSE; similarly, at larger radii ($\gtrsim 35$~kpc), the contribution from the GSE becomes small with respect to the Sagittarius dSph~\citep{Naidu20}.} of the Galactic center \citep{Deason15,Deason18,Necib19a,Naidu20}. Previous theoretical studies---e.g. \citet{Deason16,DSouza18,Elias20,Fattahi20,Santistevan20}---indicate that this may be a generic prediction of structure formation in the CDM paradigm, but the large volume and high resolution of the \TNG{} simulation allow for a new look at the subject with unprecedented statistical power. 

The panels of \autoref{fig:1} show the \exs{} stellar halo composition for two example MW analogs in the \TNG{} set, as built from the tagging procedure of \autoref{sec:2.3}. Each of the five largest mergers that contribute to the halo is indicated by a different color, ranging from the blue merger, which contributes the most stellar mass, to the gold merger, which contributes the fifth most. The contribution from all remaining mergers is shown in gray. A dashed black line indicates 50\%, the threshold used later to select significant mergers from the stellar halo. The halo shown in the left panel is overwhelmingly built of the debris of a single merger at radii $\lesssim 40$~kpc. The stellar halo shown in the right panel, however, has two quite significant mergers that together comprise half of the accreted stellar population at all radii $\lesssim 15$~kpc.

\begin{figure*}
  \centering 
  \hspace{-3.5em}
  \includegraphics{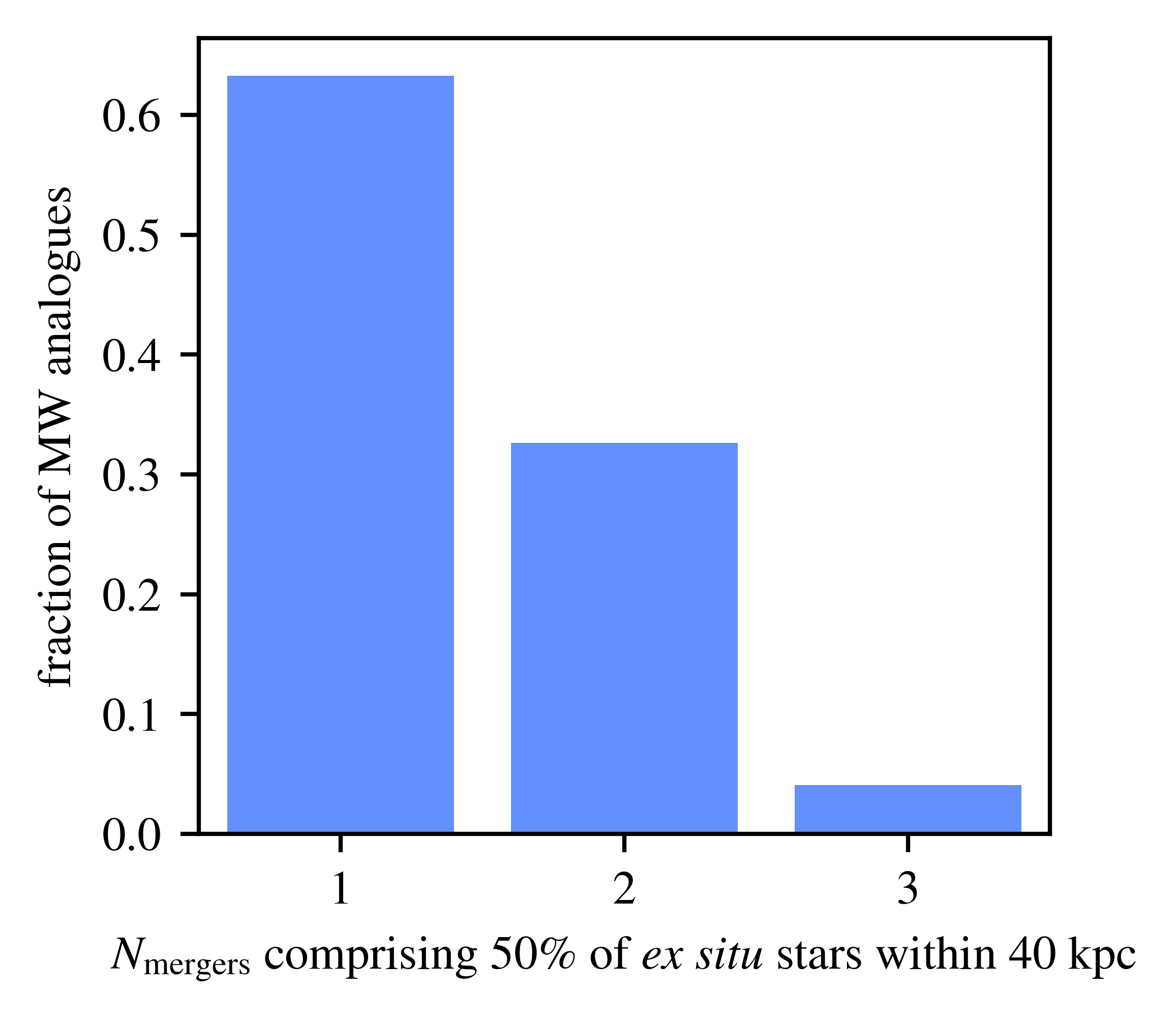}
  \includegraphics{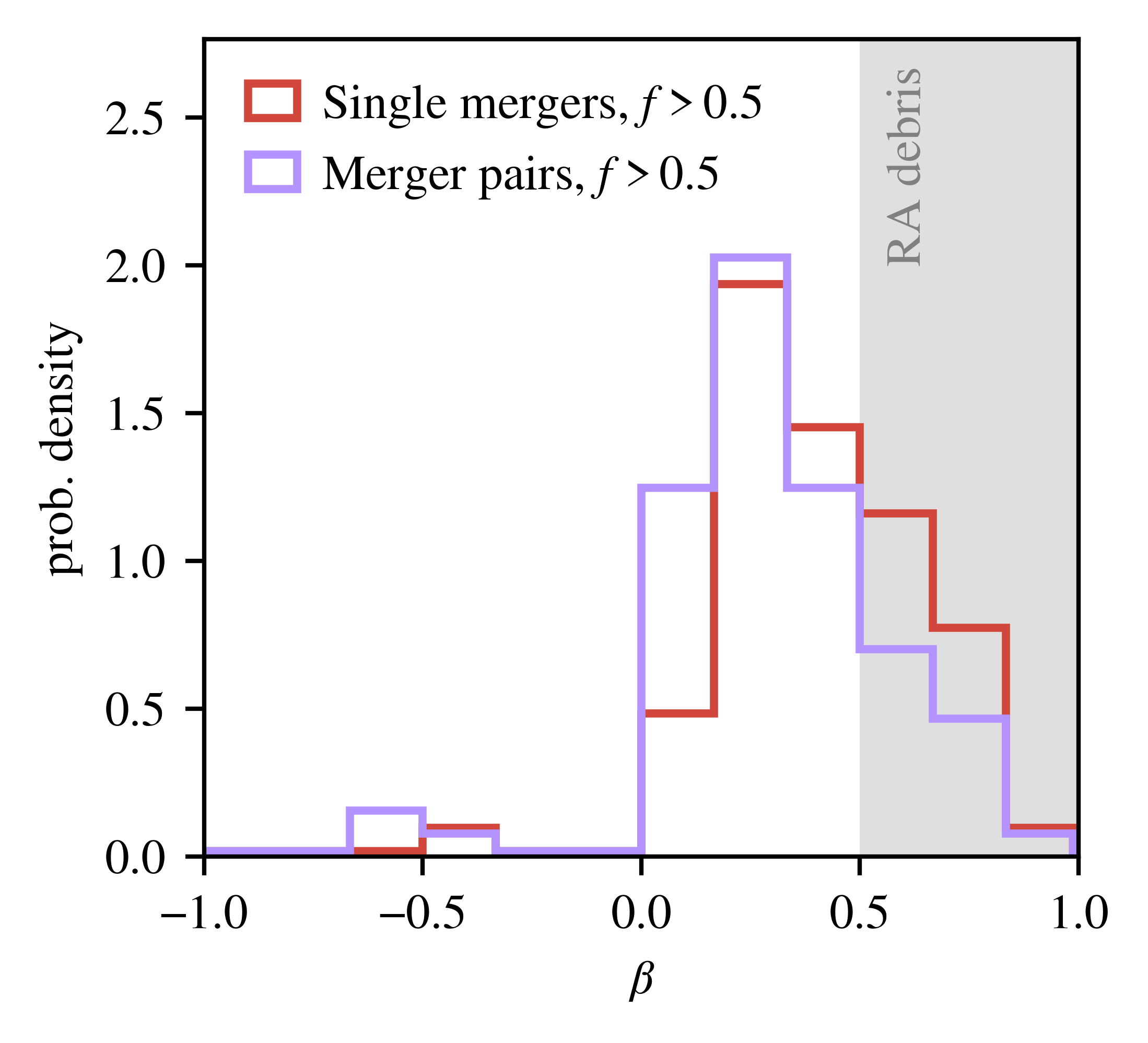}
  \caption{(\emph{Left:})~The number of mergers required to reach the threshold of 50\% of \exs{} stars within 40~kpc of the galactic center. Approximately 60\% of the MW analogs have a single merger that contributes over half of the \exs{} stars within 40~kpc, and an additional $\sim 30\%$ reach this threshold with two mergers. It is therefore not unreasonable to expect that an \exs{} halo could be predominantly comprised of two mergers, though typically there is a single large merger that comprises the majority of the inner halo. (\emph{Right:})~The velocity anisotropy $\beta$ of \exs{} stellar debris is shown for all large mergers (those that contribute over 50\% of their host's \exs{} stellar population within 40~kpc) in red. Those that are considered RA debris lie in the shaded large-$\beta$ region. Each pair of mergers that together comprise a fraction $f > 50\%$ is shown in purple. There are occasionally multiple choices of merger pairs in the same halo that satisfy the RA debris criteria (i.e., that fall in the shaded region). There are 21 single-halo RA mergers, and there are 16 RA pairs of mergers selected from 11 MW analogs (five MW analogs have multiple choices of RA merger pairs).}
  \label{fig:2}
\end{figure*}

For most stellar halos in the TNG sample~(63\%), over half of the accreted stars within 40~kpc come from a single merger. For most other halos~(33\%), summing contributions from two mergers is sufficient to reach the 50\% threshold, and for a minority~(4\%) of the halos it takes three mergers to reach this threshold, with no halos requiring four or more mergers. These statistics are summarized in the left panel of \autoref{fig:2}, which shows the fraction of MWs for which 50\% of the stellar halo is comprised of one, two, and three mergers. It is therefore not rare for an MW analog's halo to be comprised predominantly of a single merger, but it is also not rare for \emph{two} mergers to together comprise the majority of its \exs{} population. In this work, we consider GSE analogs from both of these cases. 

\subsection{GSE Analog Selection}
\label{sec:3.2}
\nopagebreak
Of the mergers identified in \TNG{}, we define a GSE-like subsample of merger debris that 
\begin{enumerate}
  \item comprises a total fraction $f > 0.5$ of all \exs{} stars within 40~kpc of the galactic center and
  \item has a velocity anisotropy $\beta > 0.5$,
\end{enumerate}
where the anisotropy is defined as the following combinations of the spherical velocity dispersions:
\begin{equation}
  \beta = 1 - \frac{\sigma^2_\phi + \sigma_\theta^2}{2 \sigma_r^2}\, .
\end{equation}
An individual merger that satisfies both criteria is deemed an ``RA merger.'' In the halos where no single merger yields $f > 0.5$ (regardless of its anisotropy), pairs of mergers are considered. In these cases, any two mergers whose stars together satisfy $f > 0.5$ and $\beta > 0.5$ are deemed an ``RA pair.'' Stars that are from either an RA merger or an RA pair are referred to as ``RA debris.''

These criteria are generally influenced by considerations of the GSE debris in our Galaxy, as the GSE comprises the majority of \exs{} material in the inner halo and has a high velocity anisotropy. Criterion~1 ensures that the GSE analog contributes significantly to the inner halo, with no other large contributions, as inferred observationally. Criterion~2 ensures that the debris qualitatively resembles the GSE kinematically, though this criterion is especially permissive. 

Of the 98 MW analogs, 21 have a single RA merger and 11 have an RA pair.\footnote{We have verified that the RA pairs are not accreted together as part of a larger group structure.} The choice of RA pair is not necessarily unique: in five halos, there are two combinations that each satisfy the requirements of an RA pair. Each combination is taken independently, though there is consistently one merger shared between the two pairs. All pairs will be specially marked in the figures, with icons denoting the MW analogs to which they belong. 

The right panel of \autoref{fig:2} presents the results of these selection criteria: the horizontal axis shows the velocity anisotropy of the stars contributed by each merger, for single and pairs of mergers with $f > 0.5$~(in red and purple, respectively). The number of halos with anisotropic \exs{} components is generally consistent with what is found in other work; population studies predict GSE-like debris in roughly a third of MW-like galaxies \citep{Fattahi19,Bose20,Elias20,Buch24}.

Notably, individual mergers tend not to have as extreme an anisotropy as the MW stellar halo; the GSE debris in the MW has $\beta \sim 0.8$ \citep{Belokurov18,Myeong18,Lancaster19,Iorio21}, which is at the far upper end of what is seen in \TNG{}. Of the debris with $f > 0.5$ (i.e., the mergers shown in the right panel of \autoref{fig:2}), 5\% of single mergers and only 2.5\% of merger pairs have $\beta > 0.8$. This implies the high anisotropy seen in the MW is quite rare.

Considering two mergers together as one set of debris tends to reduce the measured anisotropy, as the mergers do not necessarily have any correlation in velocity space; this is reflected in the shift toward isotropy ($\beta = 0$) in the right panel of \autoref{fig:2}. Despite this, there are still many pairs of mergers for which the contributed stars satisfy both the high-$\beta$ and high-$f$ requirements. This suggests that the presence of such an anisotropic structure cannot immediately preclude the possibility of a multicomponent model. The rest of this work characterizes and aims to distinguish individual GSE-like RA mergers from RA pairs.

\begin{table*}
  \setlength{\baselineskip}{4pt}\selectfont
  \renewcommand{\arraystretch}{0.8}
  \caption{Properties of the RA Mergers and RA Pairs}
  \begin{minipage}{\textwidth}
  \centering
  \begin{tabular}{@{\hspace{0em}}r@{\hspace{0em}}cr@{\hspace{0.5em}}lc*{5}{r@{\hspace{0.5em}}l}l}
  % \begin{tabular}{@{\hspace{0em}}r@{\hspace{0em}}cr@{\hspace{0.5em}}lc*{5}{r@{\hspace{0.5em}}l}l}
  % \begin{tabular}{@{\hspace{0em}}r@{\hspace{0em}}cc@{\hspace{0.5em}}cc*{5}{c@{\hspace{0.5em}}c}c}
  \toprule
  & MW ID & \multicolumn{2}{c}{$f$} & $\beta$ & \multicolumn{2}{c}{\tin{} [Gyr ago]} & \multicolumn{2}{c}{$t_{90}$ [Gyr ago]} & \multicolumn{2}{c}{$\log(\Mdyn/\Msun)$} & \multicolumn{2}{c}{$\log(\Mstar/\Msun)$} & \multicolumn{2}{c}{$\Mdyn/\mathrm{MW}~\Mdyn$} & Notes\\
  \midrule
  & 419618 & \multicolumn{2}{c}{0.96} & 0.67 & \multicolumn{2}{c}{5.87}  & \multicolumn{2}{c}{2.03} & \multicolumn{2}{c}{11.30} & \multicolumn{2}{c}{10.02} & \multicolumn{2}{c}{0.09} & \\
  & 454172 & \multicolumn{2}{c}{0.52} & 0.65 & \multicolumn{2}{c}{9.77}  & \multicolumn{2}{c}{9.58} & \multicolumn{2}{c}{11.00} & \multicolumn{2}{c}{9.37}  & \multicolumn{2}{c}{0.14} & (1) (3) \\
  & 459558 & \multicolumn{2}{c}{0.60} & 0.66 & \multicolumn{2}{c}{9.15}  & \multicolumn{2}{c}{7.75} & \multicolumn{2}{c}{11.26} & \multicolumn{2}{c}{9.47}  & \multicolumn{2}{c}{0.32} & (1) (3) \\
  & 504559 & \multicolumn{2}{c}{0.80} & 0.68 & \multicolumn{2}{c}{8.51}  & \multicolumn{2}{c}{7.69} & \multicolumn{2}{c}{11.14} & \multicolumn{2}{c}{9.71}  & \multicolumn{2}{c}{0.13} & \\
  & 505100 & \multicolumn{2}{c}{0.93} & 0.58 & \multicolumn{2}{c}{4.90}  & \multicolumn{2}{c}{3.92} & \multicolumn{2}{c}{11.29} & \multicolumn{2}{c}{10.16} & \multicolumn{2}{c}{0.23} & \\
  & 512425 & \multicolumn{2}{c}{0.58} & 0.63 & \multicolumn{2}{c}{5.87}  & \multicolumn{2}{c}{4.70} & \multicolumn{2}{c}{11.35} & \multicolumn{2}{c}{9.85}  & \multicolumn{2}{c}{0.21} & \\
  & 515296 & \multicolumn{2}{c}{0.53} & 0.82 & \multicolumn{2}{c}{7.45}  & \multicolumn{2}{c}{6.29} & \multicolumn{2}{c}{10.95} & \multicolumn{2}{c}{9.26}  & \multicolumn{2}{c}{0.07} & (4) \\
  & 520885 & \multicolumn{2}{c}{0.66} & 0.59 & \multicolumn{2}{c}{6.35}  & \multicolumn{2}{c}{2.74} & \multicolumn{2}{c}{11.01} & \multicolumn{2}{c}{9.26}  & \multicolumn{2}{c}{0.12} & \\
  & 525002 & \multicolumn{2}{c}{0.88} & 0.65 & \multicolumn{2}{c}{1.85}  & \multicolumn{2}{c}{1.34} & \multicolumn{2}{c}{11.34} & \multicolumn{2}{c}{10.04} & \multicolumn{2}{c}{0.26} & \\
  & 528322 & \multicolumn{2}{c}{0.90} & 0.58 & \multicolumn{2}{c}{4.41}  & \multicolumn{2}{c}{4.26} & \multicolumn{2}{c}{11.26} & \multicolumn{2}{c}{9.96}  & \multicolumn{2}{c}{0.29} & (2) \\
  & 530330 & \multicolumn{2}{c}{0.65} & 0.89 & \multicolumn{2}{c}{3.79}  & \multicolumn{2}{c}{4.09} & \multicolumn{2}{c}{10.95} & \multicolumn{2}{c}{9.59}  & \multicolumn{2}{c}{0.07} & \\
  & 537488 & \multicolumn{2}{c}{0.84} & 0.59 & \multicolumn{2}{c}{3.79}  & \multicolumn{2}{c}{1.01} & \multicolumn{2}{c}{11.03} & \multicolumn{2}{c}{9.29}  & \multicolumn{2}{c}{0.14} & \\
  & 537941 & \multicolumn{2}{c}{0.69} & 0.68 & \multicolumn{2}{c}{9.51}  & \multicolumn{2}{c}{8.87} & \multicolumn{2}{c}{10.68} & \multicolumn{2}{c}{8.88}  & \multicolumn{2}{c}{0.10} & \\
  & 540082 & \multicolumn{2}{c}{0.69} & 0.64 & \multicolumn{2}{c}{9.15}  & \multicolumn{2}{c}{8.11} & \multicolumn{2}{c}{10.99} & \multicolumn{2}{c}{9.62}  & \multicolumn{2}{c}{0.17} & \\
  & 540920 & \multicolumn{2}{c}{0.98} & 0.65 & \multicolumn{2}{c}{5.72}  & \multicolumn{2}{c}{7.14} & \multicolumn{2}{c}{11.38} & \multicolumn{2}{c}{10.16} & \multicolumn{2}{c}{0.28} & \\
  & 541218 & \multicolumn{2}{c}{0.92} & 0.78 & \multicolumn{2}{c}{5.87}  & \multicolumn{2}{c}{5.15} & \multicolumn{2}{c}{11.26} & \multicolumn{2}{c}{9.79}  & \multicolumn{2}{c}{0.19} & \\
  & 543729 & \multicolumn{2}{c}{0.55} & 0.77 & \multicolumn{2}{c}{3.79}  & \multicolumn{2}{c}{4.25} & \multicolumn{2}{c}{10.91} & \multicolumn{2}{c}{9.55}  & \multicolumn{2}{c}{0.11} & \\
  & 550149 & \multicolumn{2}{c}{0.51} & 0.81 & \multicolumn{2}{c}{5.37}  & \multicolumn{2}{c}{4.62} & \multicolumn{2}{c}{10.96} & \multicolumn{2}{c}{9.46}  & \multicolumn{2}{c}{0.12} & \\
  & 552414 & \multicolumn{2}{c}{0.68} & 0.73 & \multicolumn{2}{c}{8.51}  & \multicolumn{2}{c}{8.62} & \multicolumn{2}{c}{10.91} & \multicolumn{2}{c}{9.38}  & \multicolumn{2}{c}{0.13} & \\
  & 569251 & \multicolumn{2}{c}{0.83} & 0.60 & \multicolumn{2}{c}{10.06} & \multicolumn{2}{c}{7.07} & \multicolumn{2}{c}{10.35} & \multicolumn{2}{c}{8.55}  & \multicolumn{2}{c}{0.05} & \\
  & 571633 & \multicolumn{2}{c}{0.69} & 0.60 & \multicolumn{2}{c}{7.45}  & \multicolumn{2}{c}{7.02} & \multicolumn{2}{c}{10.77} & \multicolumn{2}{c}{9.38}  & \multicolumn{2}{c}{0.14} & \\
  \midrule[\heavyrulewidth]
  & MW ID & \multicolumn{2}{c}{$f$} & $\beta_\mathrm{tot}$ & \multicolumn{2}{c}{\tin{} [Gyr ago]}& \multicolumn{2}{c}{$t_{90}$ [Gyr ago]} & \multicolumn{2}{c}{$\log(\Mdyn/\Msun)$} & \multicolumn{2}{c}{$\log(\Mstar/\Msun)$} & \multicolumn{2}{c}{$\Mdyn/\mathrm{MW}~\Mdyn$} & Notes\\
  \cmidrule(lr){3-4}\cmidrule(lr){6-7}\cmidrule(lr){8-9}\cmidrule(lr){10-11}\cmidrule(lr){12-13}\cmidrule(lr){14-15}
  &  & High $f$ & Low $f$ &  & High $f$ & Low $f$ & High $f$ & Low $f$ & High $f$ & Low $f$ & High $f$ & Low $f$ & High $f$ & Low $f$ & \\
  \midrule
  \ldelim\{{2}{1.5ex} & 392277 & 0.39 & 0.20 & 0.85 & 3.27  & 10.82 & 2.60  & 10.10 & 10.73 & 10.65 & 8.95 & 8.35 & 0.05 & 0.12 & (1) (3) (4) (5) \\
                      & 392277 & 0.39 & 0.19 & 0.83 & 3.27  & 11.86 & 2.60  & 11.94 & 10.73 & 10.65 & 8.95 & 8.29 & 0.05 & 0.31 & (1) (3) (4) (5) \\
                      & 394621 & 0.28 & 0.28 & 0.61 & 7.28  & 10.82 & 6.35  & 10.26 & 10.93 & 10.66 & 9.12 & 8.79 & 0.07 & 0.07 & (2) (4) (5) \\
                      & 446665 & 0.40 & 0.21 & 0.63 & 9.30  & 7.92  & 8.80  & 6.90  & 11.01 & 10.89 & 9.58 & 9.44 & 0.13 & 0.06 & (5) \\
                      & 459557 & 0.41 & 0.27 & 0.54 & 11.12 & 11.99 & 10.65 & 11.92 & 10.20 & 9.96  & 8.06 & 7.89 & 0.05 & 0.05 & (3) (5) \\
  \ldelim\{{2}{1.5ex} & 495451 & 0.37 & 0.16 & 0.69 & 10.36 & 11.86 & 9.92  & 11.61 & 10.73 & 10.67 & 9.14 & 8.65 & 0.07 & 0.12 & \\
                      & 495451 & 0.37 & 0.23 & 0.80 & 10.36 & 6.98  & 9.92  & 6.21  & 10.73 & 10.96 & 9.14 & 9.02 & 0.07 & 0.07 & \\
                      & 516760 & 0.31 & 0.24 & 0.57 & 12.26 & 11.42 & 12.11 & 11.12 & 10.61 & 10.60 & 8.76 & 8.65 & 0.18 & 0.07 & \\
  \ldelim\{{2}{1.5ex} & 534628 & 0.29 & 0.21 & 0.70 & 6.67  & 10.06 & 5.33  & 9.60  & 10.94 & 10.55 & 9.15 & 8.99 & 0.09 & 0.08 & \\
                      & 534628 & 0.40 & 0.29 & 0.66 & 9.51  & 6.67  & 8.85  & 5.33  & 10.78 & 10.94 & 9.28 & 9.15 & 0.11 & 0.09 & \\
  \ldelim\{{2}{1.5ex} & 535050 & 0.46 & 0.08 & 0.60 & 11.86 & 12.84 & 11.36 & 12.54 & 10.17 & 9.70  & 8.10 & 7.34 & 0.08 & 0.15 & (5) \\
                      & 535050 & 0.46 & 0.07 & 0.62 & 11.86 & 12.26 & 11.36 & 11.85 & 10.17 & 10.06 & 8.10 & 7.37 & 0.08 & 0.10 & (5) \\
                      & 538905 & 0.31 & 0.29 & 0.69 & 6.98  & 10.67 & 6.92  & 10.29 & 10.75 & 10.65 & 9.23 & 8.97 & 0.07 & 0.18 & \\
                      & 543114 & 0.39 & 0.34 & 0.53 & 11.66 & 11.66 & 11.43 & 10.34 & 10.79 & 10.33 & 8.83 & 8.52 & 0.14 & 0.05 & \\
  \ldelim\{{2}{1.5ex} & 552581 & 0.47 & 0.07 & 0.74 & 9.30  & 12.69 & 9.17  & 12.44 & 10.81 & 9.88  & 9.27 & 8.11 & 0.12 & 0.16 & \\
                      & 552581 & 0.47 & 0.40 & 0.62 & 9.30  & 10.67 & 9.17  & 10.05 & 10.81 & 10.47 & 9.27 & 8.90 & 0.12 & 0.09 & \\
  \bottomrule
  \end{tabular}
  \end{minipage}
  \tablecomments{Various properties of the mergers that contribute RA debris to the MW analogs. The top section of the table describes single RA mergers, and the bottom section describes RA pairs. The columns are as follows: the \Subfind~ID of the MW analog that hosts the RA debris; the fraction $f$ of \exs{} stars within 40~kpc of the galactic center that are from the mergers comprising the RA debris; the velocity anisotropy $\beta$ of these stars; the infall time \tin{} of the merger; the time $t_{90}$ at which 90\% of the merging galaxy's stars had formed; the total bound mass of the galaxy; the stellar mass of the galaxy; the ratio of the bound mass of the infalling galaxy to that of the MW host; and various notes, described below. All masses are evaluated at $\tin$. The notes are as follows: (1)~the MW analog is non-central in its FoF group, (2)~the MW analog is within 100 kpc of a halo with $\Mdyn > 10^{10}~\Msun$, (3)~the MW analog is within 1~Mpc of an M31-like partner halo (i.e., a halo that has a larger $\Mdyn$), (4)~the dwarf galaxy progenitor of the RA debris survives to $z = 0$---in the case of the RA pairs, only the large-$f$ galaxy survives in these MW analogs, and (5)~the GSE-like event is not the host's most recent major merger.}
  \label{tab:2}
\end{table*}

% \Needspace*{4\baselineskip}
\subsection{Progenitors of GSE-like Debris}
\label{sec:3.3}
\nopagebreak
Various properties of GSE analogs are shown in \autoref{tab:2}, for both the RA mergers (top section) and RA pairs (bottom section). The columns include
\begin{enumerate}
	\item MW ID, the \Subfind{}~ID of the MW analog to which the GSE analog belongs;
	\item $f$, the fractional contribution of \exs{} stars in the inner 40~kpc belonging to the GSE analog;
	\item $\beta$, the velocity anisotropy of all the stars comprising the RA debris;
	\item $\tin$, the infall time of the GSE-like mergers;
	\item $t_{90}$, the time by which 90\% of the mergers' stars form;
	\item $\Mdyn$, the total mass of the mergers at $\tin$;
	\item $M_\star$, the stellar mass of the mergers at $\tin$; and 
	\item $\Mdyn/\mathrm{MW}~\Mdyn$, the ratio of the merger's mass to the MW host's mass at $\tin$. 
\end{enumerate}
The final column includes notes for cases of particular interest. For the RA pairs, most of the information is given for each contributing merger separately, where the mergers are ordered by $f$. Cases where there are multiple choices of RA pair in the same stellar halo are denoted by braces on the left edge of the table. 

\begin{figure}
  \centering
  \includegraphics{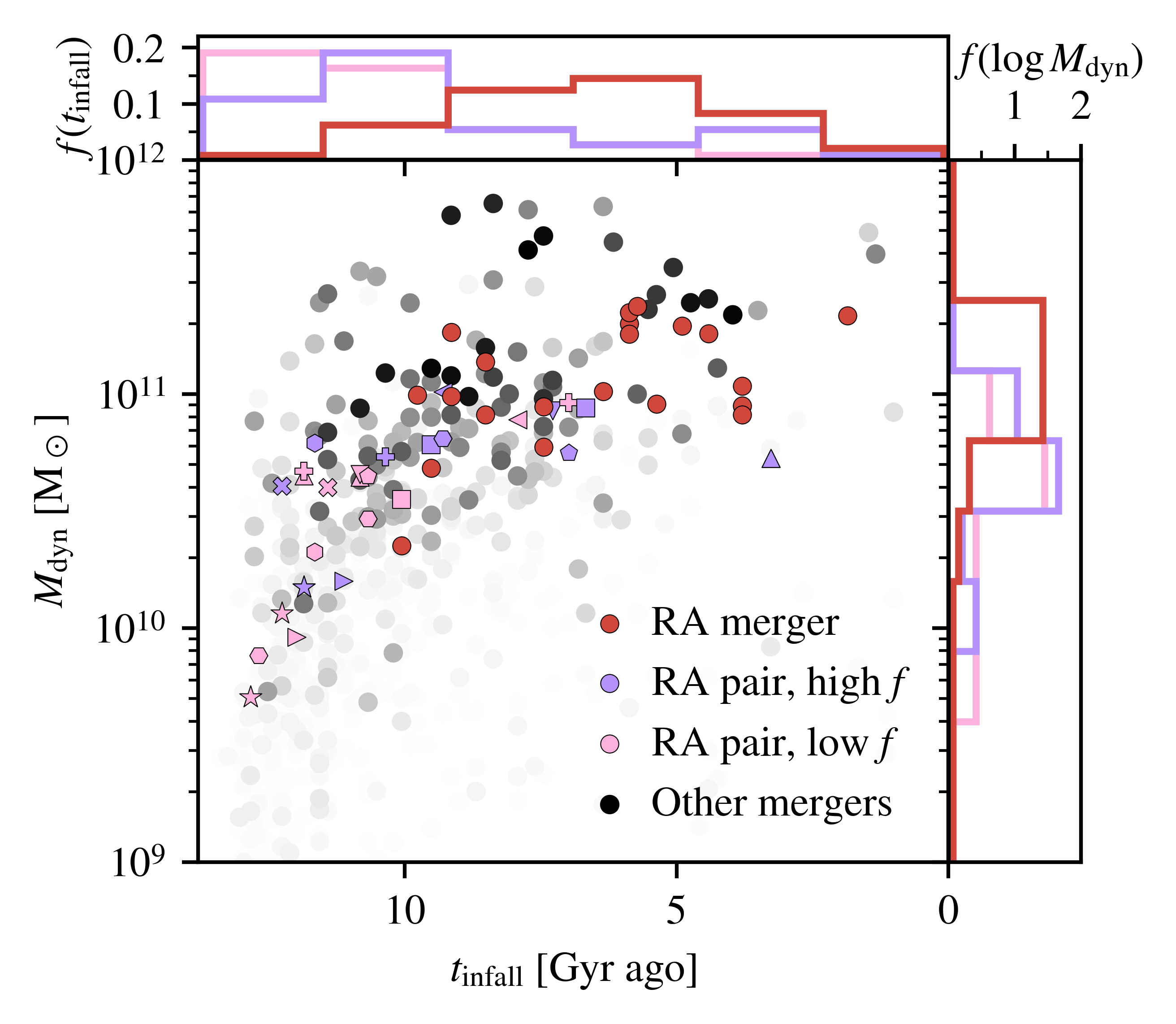}
  \caption{The infall times $\tin$ and masses $\Mdyn$ of all mergers across the TNG MW analogs. The RA mergers passing the selection criteria established in \autoref{sec:3.2} are shown in red, while RA pairs are shown in purple (for the larger-$f$ merger) and pink (for the smaller-$f$ merger), with marker shapes indicating the MW analog to which they belong. Other (non-RA) mergers are shown in black, with the opacity equal to $f$, with higher-$f$ mergers represented by darker markers. The corresponding marginal distributions are shown above and to the right of the main panel. Generally, RA mergers are more recently accreted and larger in mass than those comprising RA pairs. For the RA pairs, the larger-$f$ mergers tend to be more recently accreted and have slightly larger mass.}
  \label{fig:3}
\end{figure}

\begin{figure*}
  \centering
  \hspace{-3.5em}
  \includegraphics{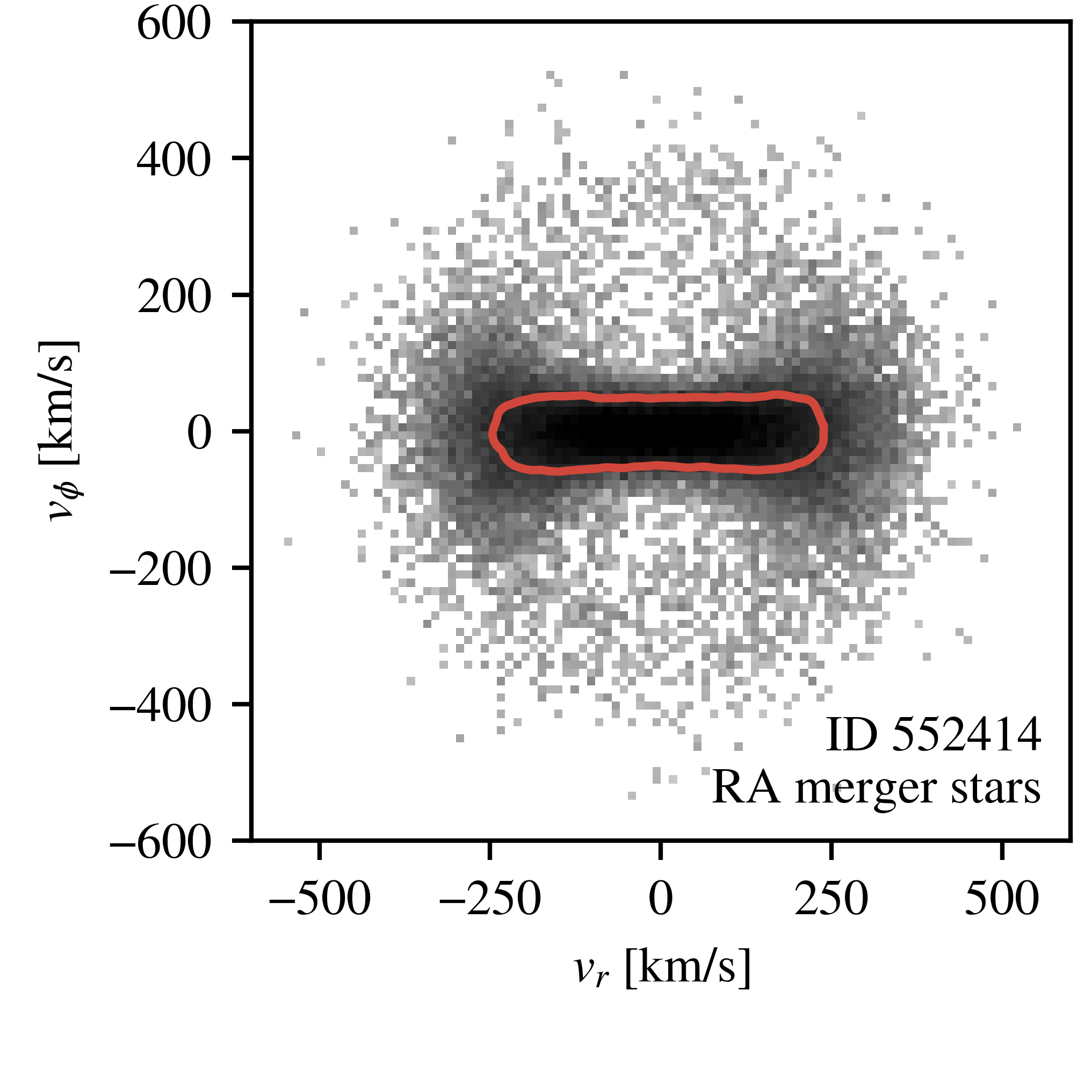}
  \includegraphics{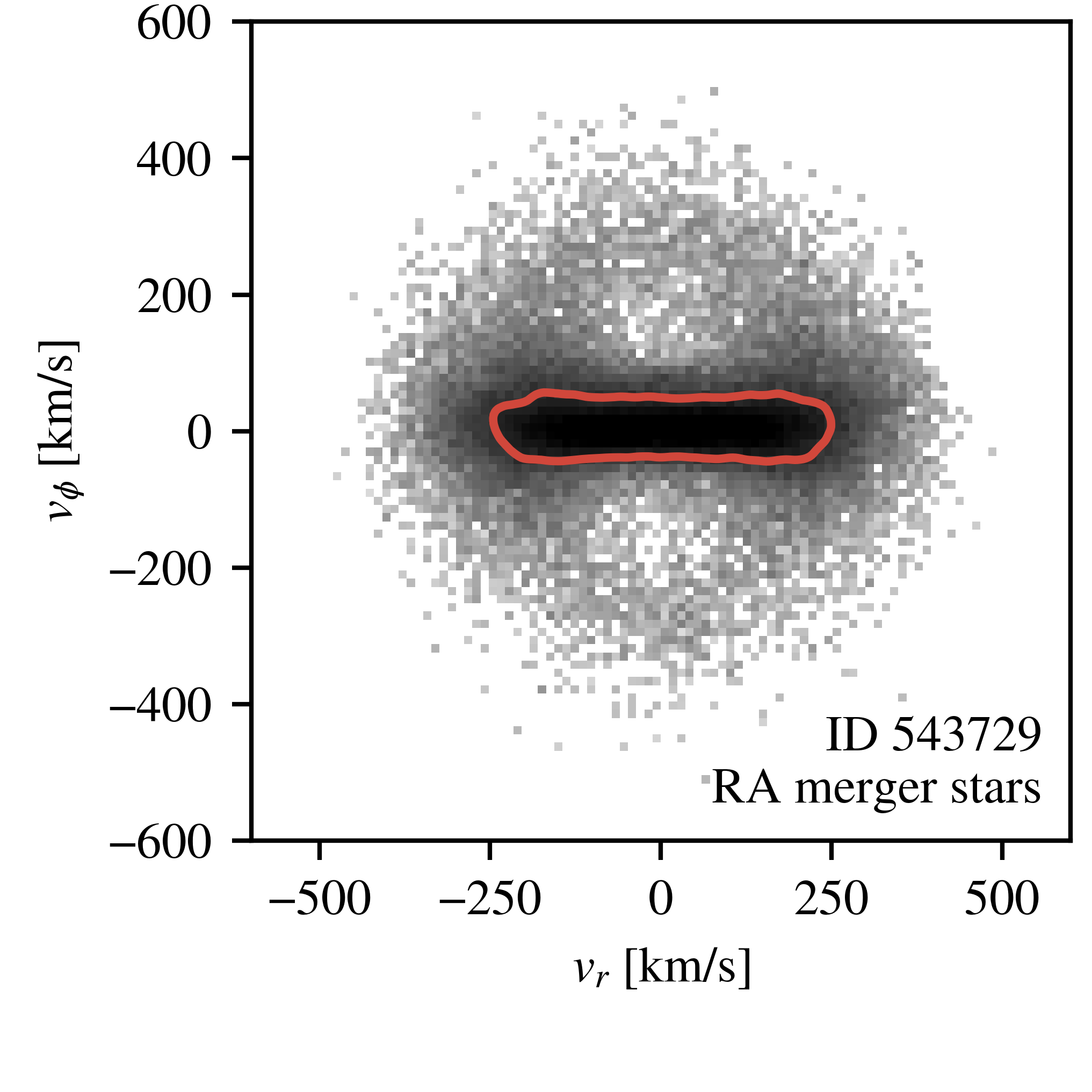}\\
  \hspace{-3.5em}
  \includegraphics{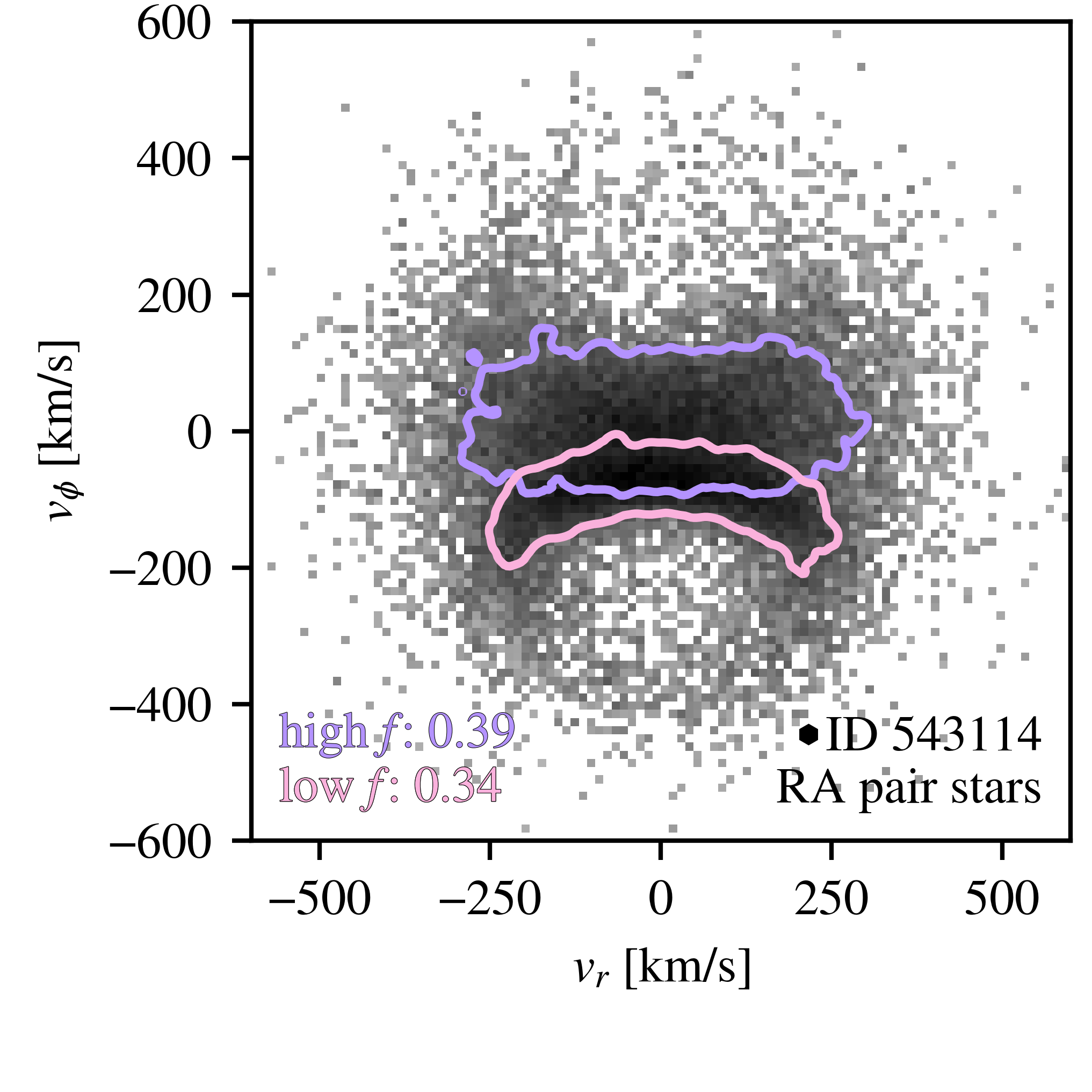}
  \includegraphics{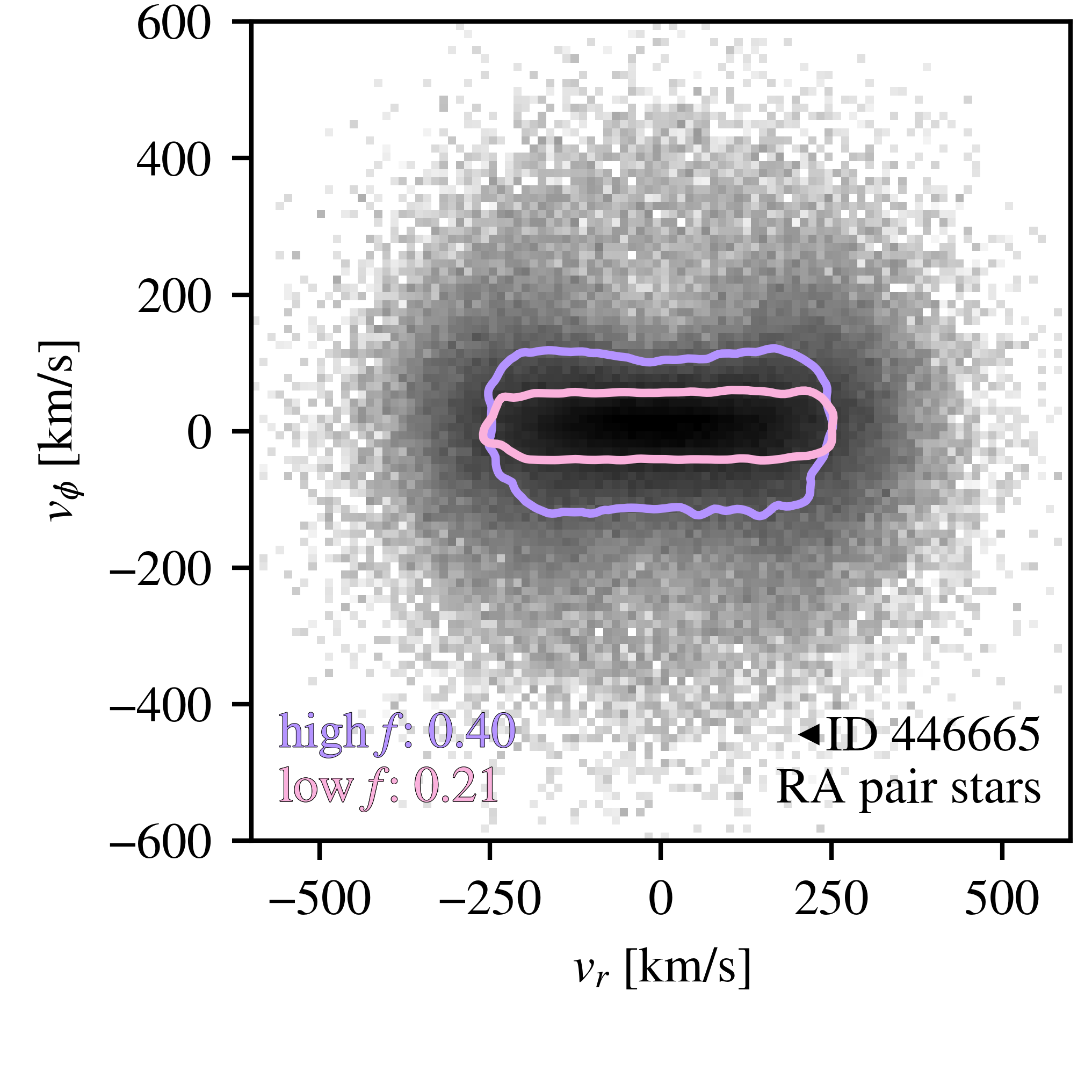}
  \caption{The distribution of spherical velocities $v_r$ and $v_\phi$ for RA merger debris in four example MW analogs. The gray-scale histogram shows the full distribution of RA merger debris. The top row shows these distributions for single RA mergers, with 68\% of the stars enclosed within the red contour. The bottom row shows the distributions for RA pairs, with 68\% of stars from the high-$f$~(low-$f$) merger enclosed within the purple~(pink) contour. In each bottom panel, the fractions $f$ are provided in the lower left corner, and an icon unique to the MW analog is provided in the lower right corner, consistent with \autoref{fig:3}. The examples chosen here highlight the diversity in kinematic features observed in MW's with RA debris. Versions of this figure for each choice of RA debris are provided in \autoref{app:C}.}
  \label{fig:4}
\end{figure*}

\begin{figure*}
  \centering
  \hspace{-3.5em}
  \includegraphics{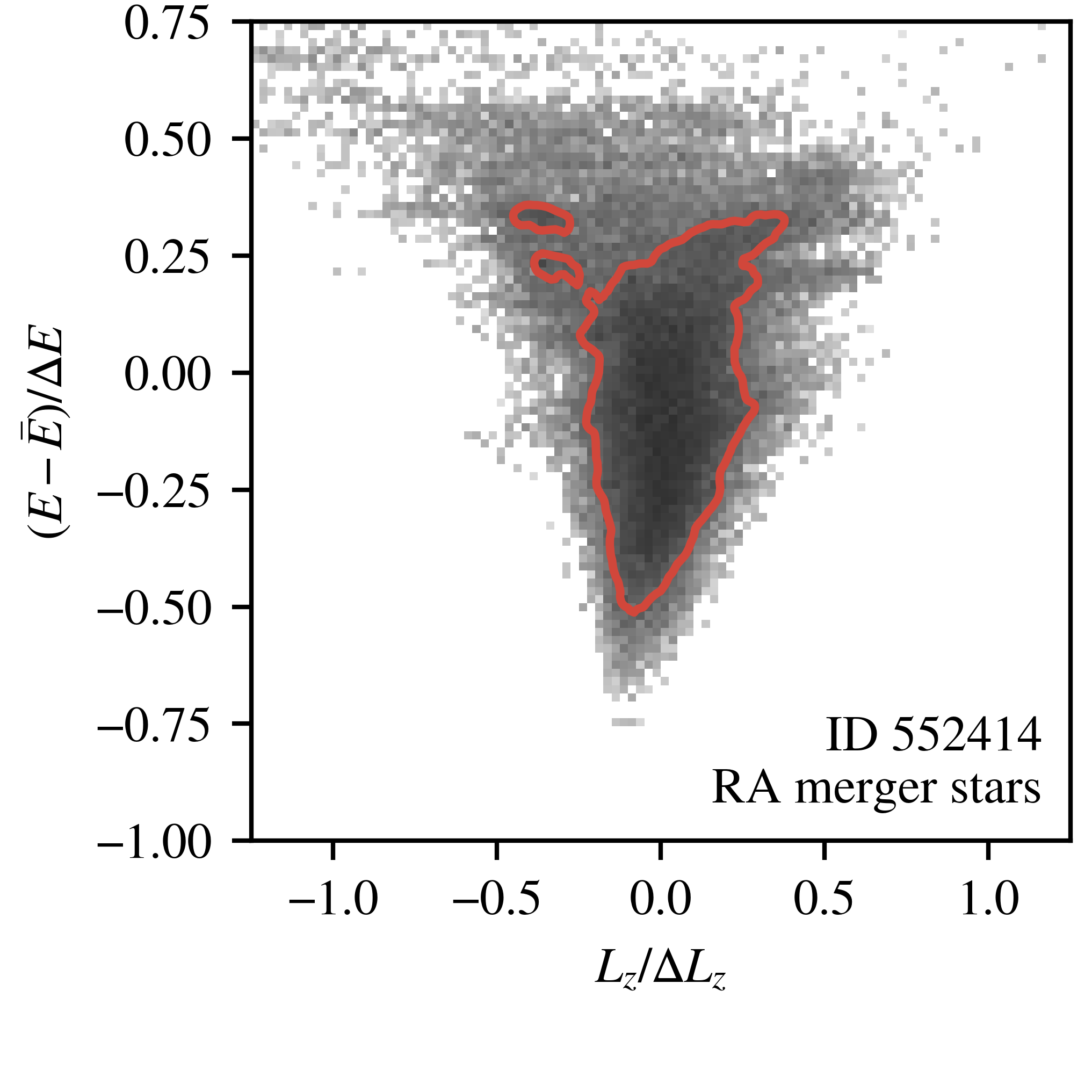}
  \includegraphics{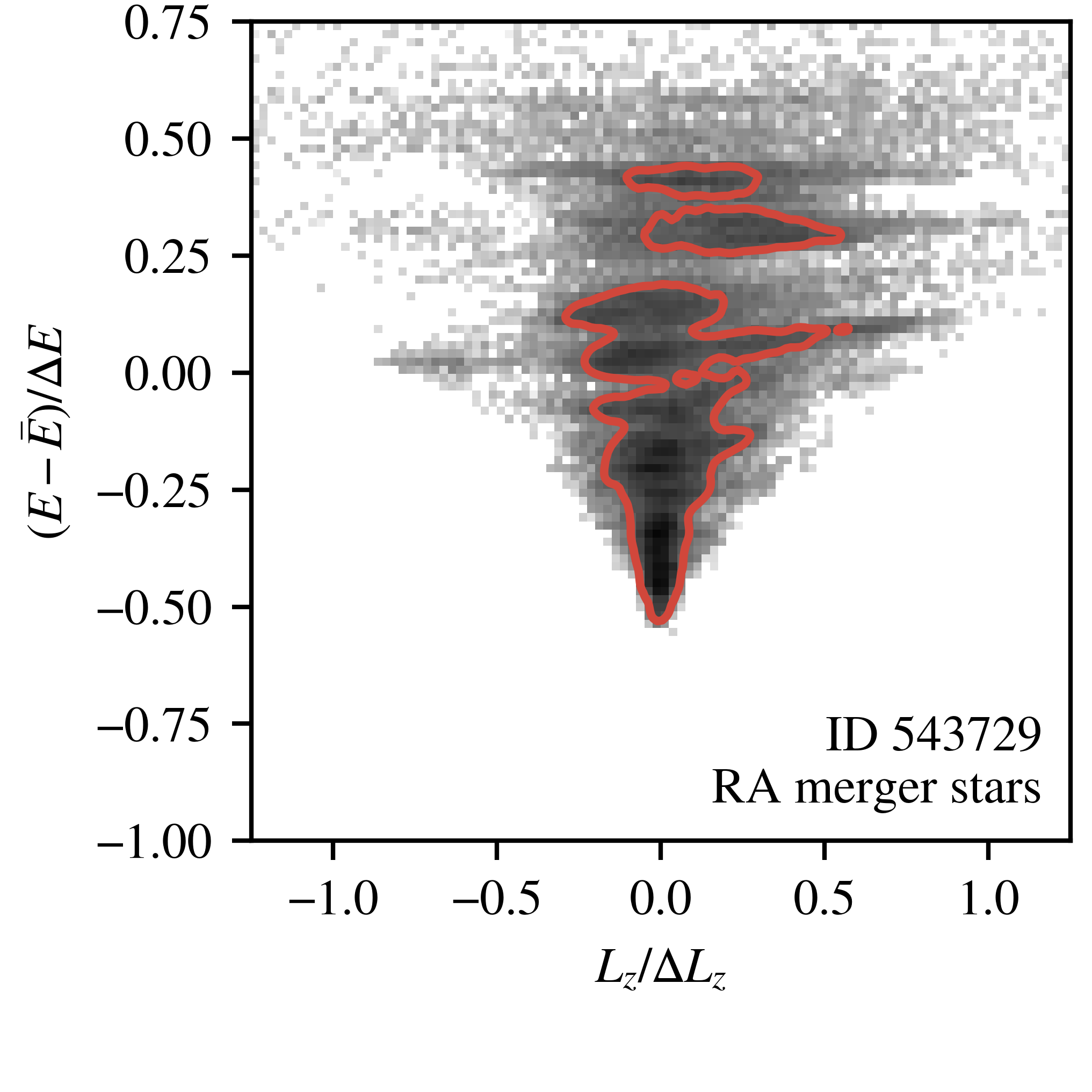}\\
  \hspace{-3.5em}
  \includegraphics{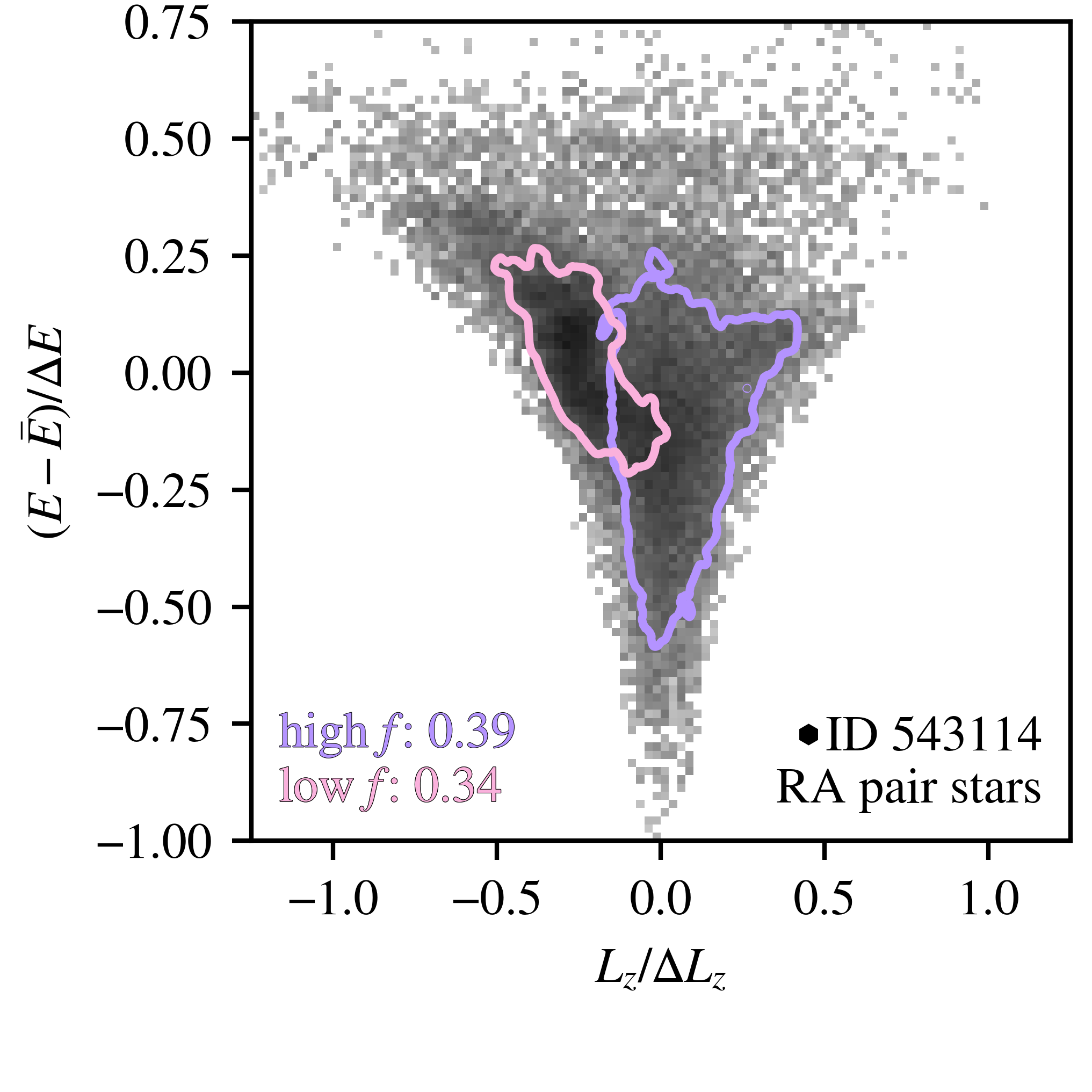}
  \includegraphics{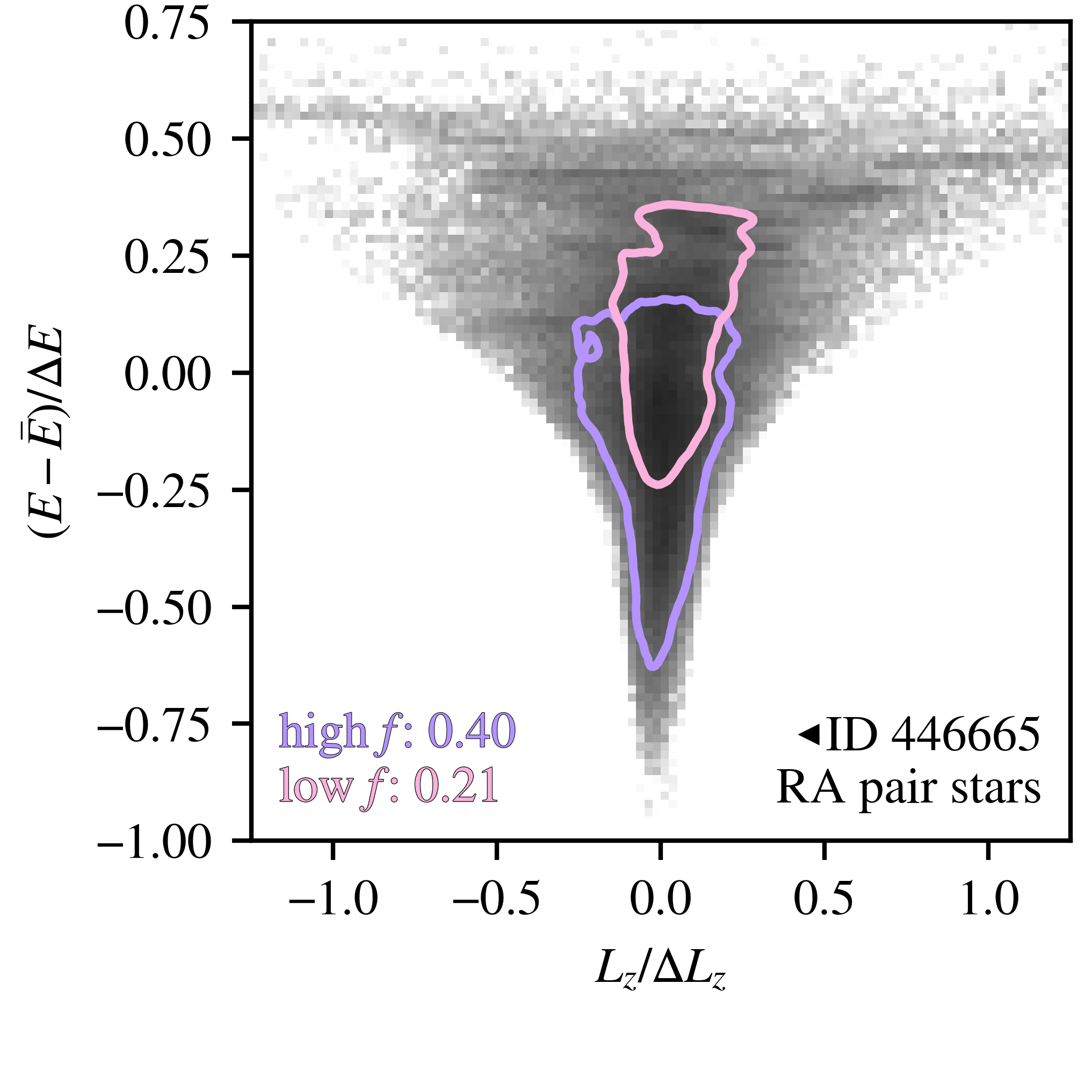}
  \caption{Total energy shown against the $z$-component of angular momentum for the same MW analogs as in \autoref{fig:4}, with identical formatting, except as noted. Each axis has been normalized to an energy and angular momentum scale characteristic of the RA debris to emphasize the shapes of these distributions. Specifically, $\bar{E}$ is the median energy of RA debris, while $\Delta E$ is the difference between the 2nd and 98th percentile. Similarly, $\Delta L_{z}$ is the difference between the 2nd and 98th percentile values of the RA debris' $L_z$. As in the $v_r - v_\phi$ plane, these distributions exhibit varying degrees of complexity regardless of whether they correspond to single or pairs of mergers. Versions of this figure for each choice of RA debris are provided in \autoref{app:C}.}
  \label{fig:5}
\end{figure*}

Across the sample, there are only three cases in which an infalling GSE-like galaxy survives to $z = 0$. This happens once for a single RA merger (in ID~515296) and twice for the larger-$f$ partner of an RA pair (in MWs 392277 and 394621). In all cases where the GSE-like debris is built up predominantly from a single merger, that event is the most recent major merger (i.e., no subsequent merger has larger $\Mdyn/\mathrm{MW}~\Mdyn$). However, there are five MWs that accrete an RA pair and subsequently have a merger that is more significant than either in the pair. This happens in IDs~392277, 394621, 446665, 459557, and 535050. 

Further elements of the table are summarized in \autoref{fig:3}, which shows the distribution of infall times, $\tin$, and total masses at infall, $\Mdyn$, for RA mergers (in red), RA pairs (in purple and pink, with the larger-$f$ merger in purple), and all other mergers (in black, with opacity equal to the fraction $f$) across the sample. The marginal distributions of these parameters for the RA debris are shown as normalized histograms on each axis.

Generally, RA mergers are more recent ($\tin=5.9_{-2.0}^{+3.3}$~Gyr ago) and larger ($\log\Mdyn/\Msun = 11.0^{+0.3}_{-0.1}$) than mergers contributing to RA pairs. Of those that contribute to RA pairs, the merger with the larger $f$ is typically more recent ($\tin = 9.4^{+2.4}_{-2.7}$ versus $11.1^{+1.1}_{-2.4}$~Gyr ago) and has a slightly larger mass ($\log\Mdyn/\Msun = 10.7^{+0.2}_{-0.4}$ versus $10.6^{+0.2}_{-0.7}$). This systematic shift in merger populations is particularly manifest in their SFHs, explored in \autoref{sec:5}, but we first characterize the kinematic properties of these GSE analogs. 

\Needspace*{4\baselineskip}
\section{Kinematic Features}
\label{sec:4}
As summarized in \autoref{tab:2}, a number of GSE-like events are identified in \TNG{}'s MW analogs. In general, RA debris is created by a single merger or---in a non-trivial fraction of cases---by pairs of mergers. A natural question that follows is whether or not these two scenarios can be distinguished using observable properties of the stellar halos. To this end, this section explores the kinematic properties of the stars in the RA mergers and pairs, such as their velocities, orbital energies and angular momenta. \autoref{sec:5} will focus instead on stellar ages and abundances.

\autoref{fig:4} shows the distribution of spherical galactocentric velocities $v_r$ versus $v_\phi$ for the RA debris of four MW analogs. The star counts are indicated by the gray-scale histogram, while the contours enclose 68\% of the stars from each merger: the single RA merger in red and the large-$f$~(small-$f$) merger of the RA pair in purple~(pink). The four panels illustrate the kinematic diversity across different MWs in the TNG set. The top row of \autoref{fig:4} provides two examples of single RA mergers. In the left panel, the debris has a clear ``sausage''-like velocity distribution, with a distinctive extension along $v_r$. In the right panel, the RA merger shows a greater population of stars on less eccentric orbits, with $v_r\sim 0$ and nonzero $v_\phi$, making the full extent of the debris potentially more difficult to distinguish from older contributions to the \exs{} stellar halo that are expected to be nearly isotropic~\citep{Necib19}. 

The bottom row of \autoref{fig:4} highlights two examples of RA pairs. On the left, the debris from both mergers is reasonably separated in velocity-space, while on the right, it is not. Quantitatively, for the left panel, 18\% of stars from the high-$f$ merger are contained within the pink contour, while 45\% of those from the low-$f$ merger fall within the purple contour. Conversely, for the right panel, 39\% of stars from the high-$f$ merger are contained within the pink contour, while 79\% of those from the low-$f$ merger fall within the purple contour. For this last scenario, it would be impossible to separate out the two contributions based on kinematics alone. 

Typically, single RA mergers have lower $v_\phi$ dispersion than RA pairs: the former have $\sigma_\phi = 86^{+13}_{-18}~\mathrm{km\ s}^{-1}$ while the latter have $\sigma_\phi = 95^{+14}_{-11}~\mathrm{km\ s}^{-1}$. However, because there is significant scatter in the azimuthal dispersion from halo to halo, the precise value of $\sigma_\phi$ does little to distinguish an individual RA merger from a pair. The other components of velocity are more comparable within their scatter. In the radial direction, single RA mergers exhibit a dispersion of $\sigma_r = 159_{-27}^{+18}~\mathrm{km\ s}^{-1}$ versus the pairs' $164_{-21}^{+29}~\mathrm{km\ s}^{-1}$. In the polar direction, RA mergers have $\sigma_\theta = 90_{-20}^{+12}~\mathrm{km\ s}^{-1}$ versus the pairs' $89_{-5}^{+12}~\mathrm{km\ s}^{-1}$. These trends (in addition to the trends in $\beta$ shown in the right panel of \autoref{fig:2}) do indicate that debris from the RA pairs is slightly more isotropized, befitting of their earlier infall times (see~\autoref{fig:3}), though this is only evident in their $v_\phi$ values.

The debris from the same MWs is shown in \autoref{fig:5}, where the distribution of orbital energy $E$ is plotted against $L_z$, the $z$-component of angular momentum.\footnote{Because the shape of the distribution is more important than the numerical values of $E$ and $L_z$, the axes are rescaled. In particular, the energy is centered at $\bar{E}$, the median energy of the RA debris, and normalized by $\Delta E$, the difference between the 2nd and 98th percentiles. Similarly, though $L_z$ is still centered at zero, it is normalized by $\Delta L_{z}$, the difference between the 2nd and 98th percentile values of the RA debris' $L_z$.} As in \autoref{fig:4}, the top row contrasts two examples of single RA mergers. The example on the right has more complex substructure, with debris deposited at distinct energies. On their own, distinct clumps in energy and angular momentum could potentially be misinterpreted as originating from multiple mergers---indeed, due to the inherent difficulties in effectively clustering stars in observational parameters, a standard approach is to accept this fact and intentionally over-fragment with clustering algorithms, so that multiple clusters of stars can later be associated to the same object~\citep[e.g.][]{Koppelman19,Yuan20,Lovdal22,Ou23}. These analyses must therefore be performed with great care, keeping in mind the potential complexities of stellar debris in these kinematic spaces. 

\begin{figure*}
  \centering
  \hspace{-3.5em}
  \includegraphics{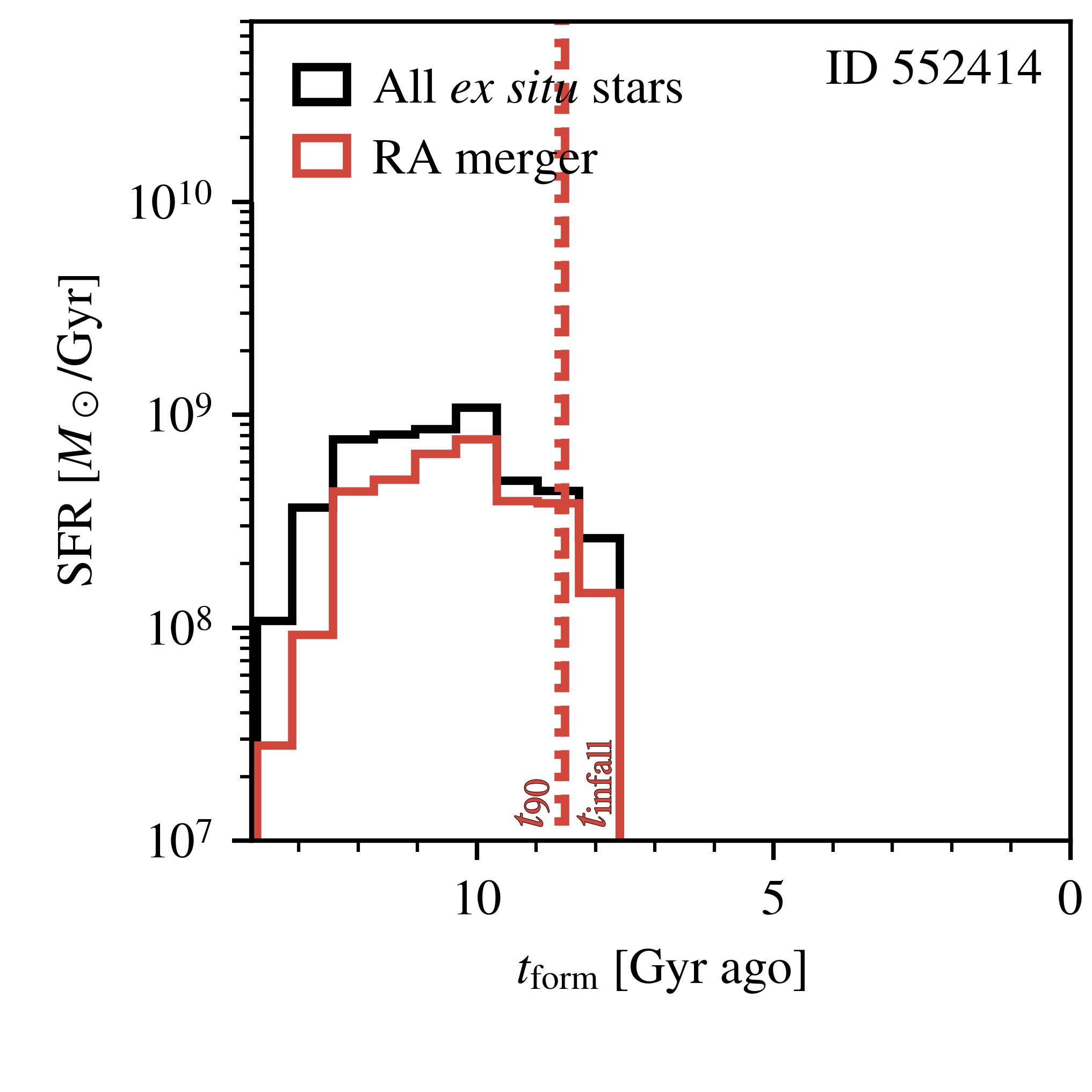}
  \includegraphics{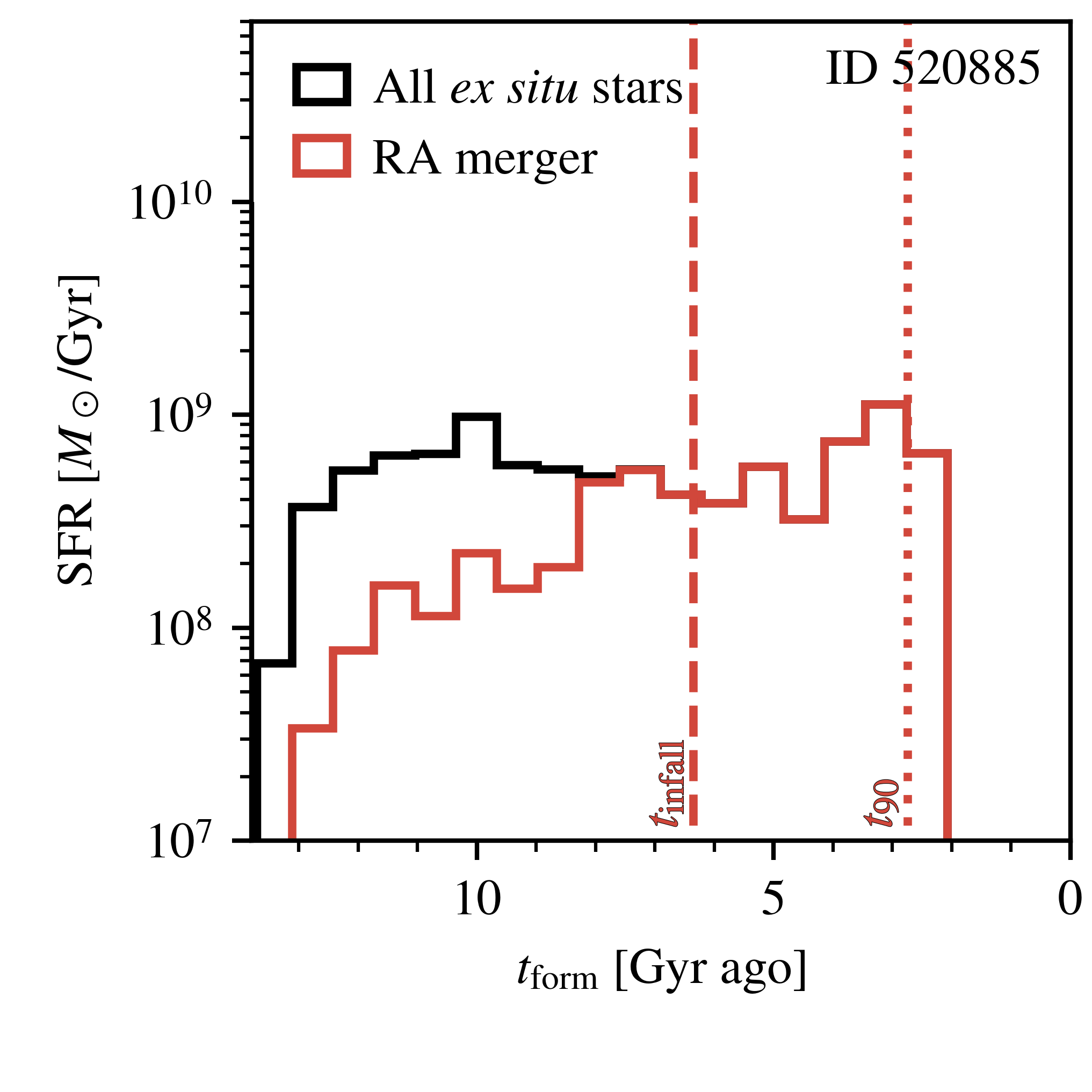}\\
  \hspace{-3.5em}
  \includegraphics{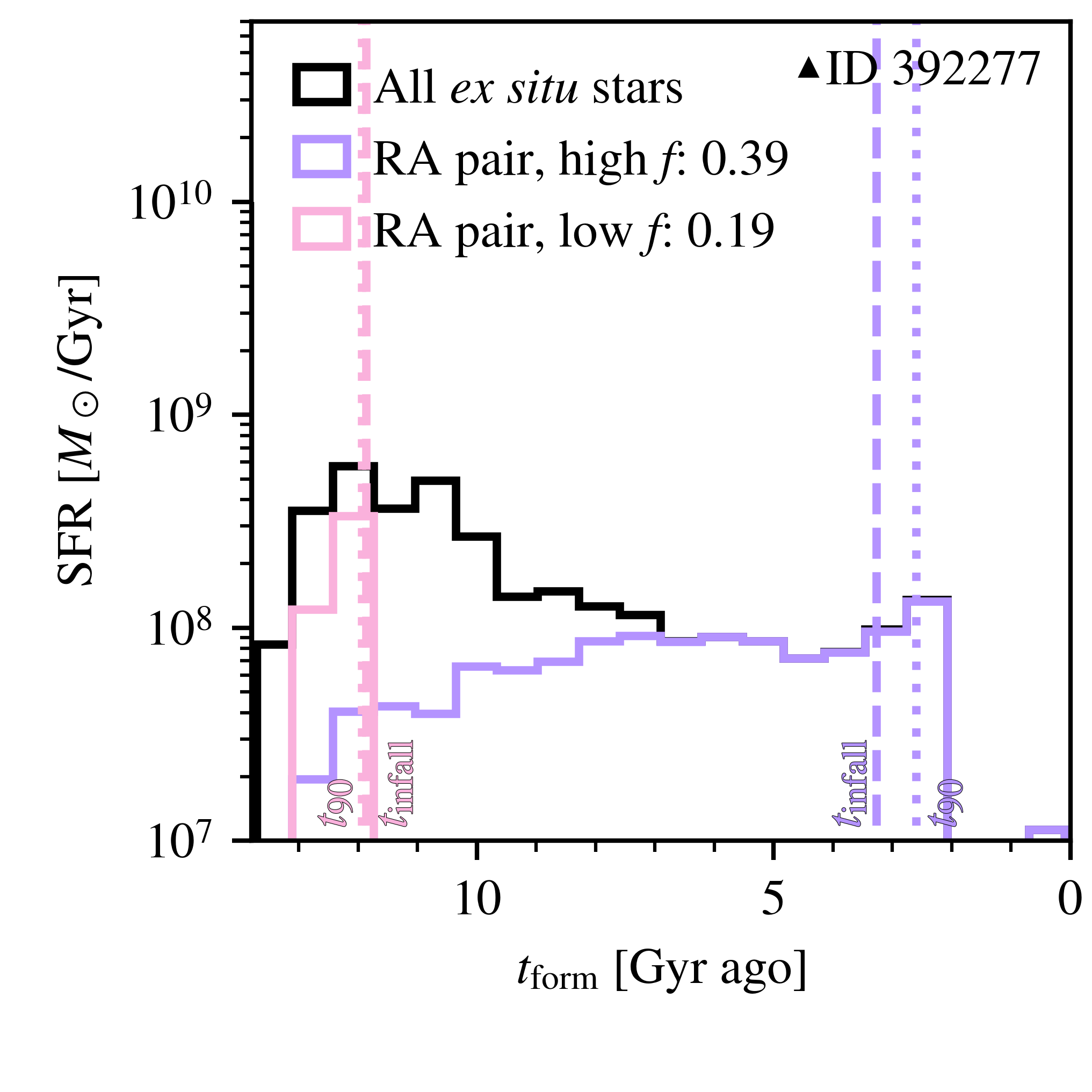}
  \includegraphics{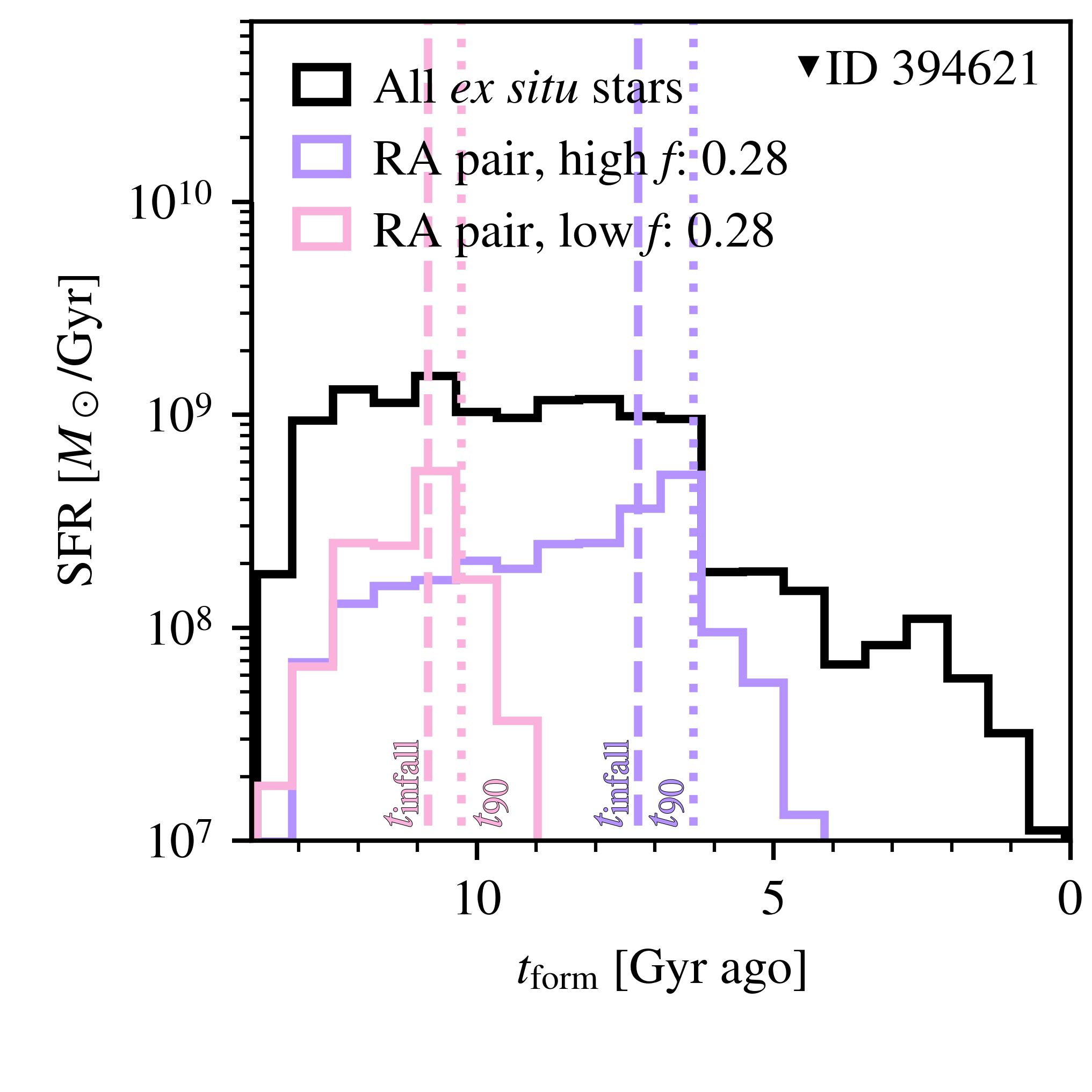}
  \caption{SFHs for the \exs{} stars of four MW analogs with RA debris. In each panel, the black histogram shows the total \exs{} star formation, while the colored lines show the contribution from individual mergers, with single RA mergers in red, large-$f$ components of RA pairs in purple, and small-$f$ components of RA pairs in pink. The vertical dashed line shows the infall time, \tin{}, of each merger, and the dotted line shows the quenching time, $t_{90}$, by which 90\% of stars have formed. The left column highlights cases in which star formation ceases shortly after the galaxies enter the virial radius of their host. In the case of the RA pair (which is given a marker consistent with previous figures), the periods of star formation of the two galaxies comprising the pair are well-separated in time, making them easier to distinguish from one another. The right column highlights examples where star formation continues long after the galaxies are accreted onto their hosts and where the RA pair does not show the separation seen in the left panel. Versions of this figure for each choice of RA debris are provided in \autoref{app:C}.}
  \label{fig:6}
\end{figure*}

\begin{figure}
  \centering
  \includegraphics{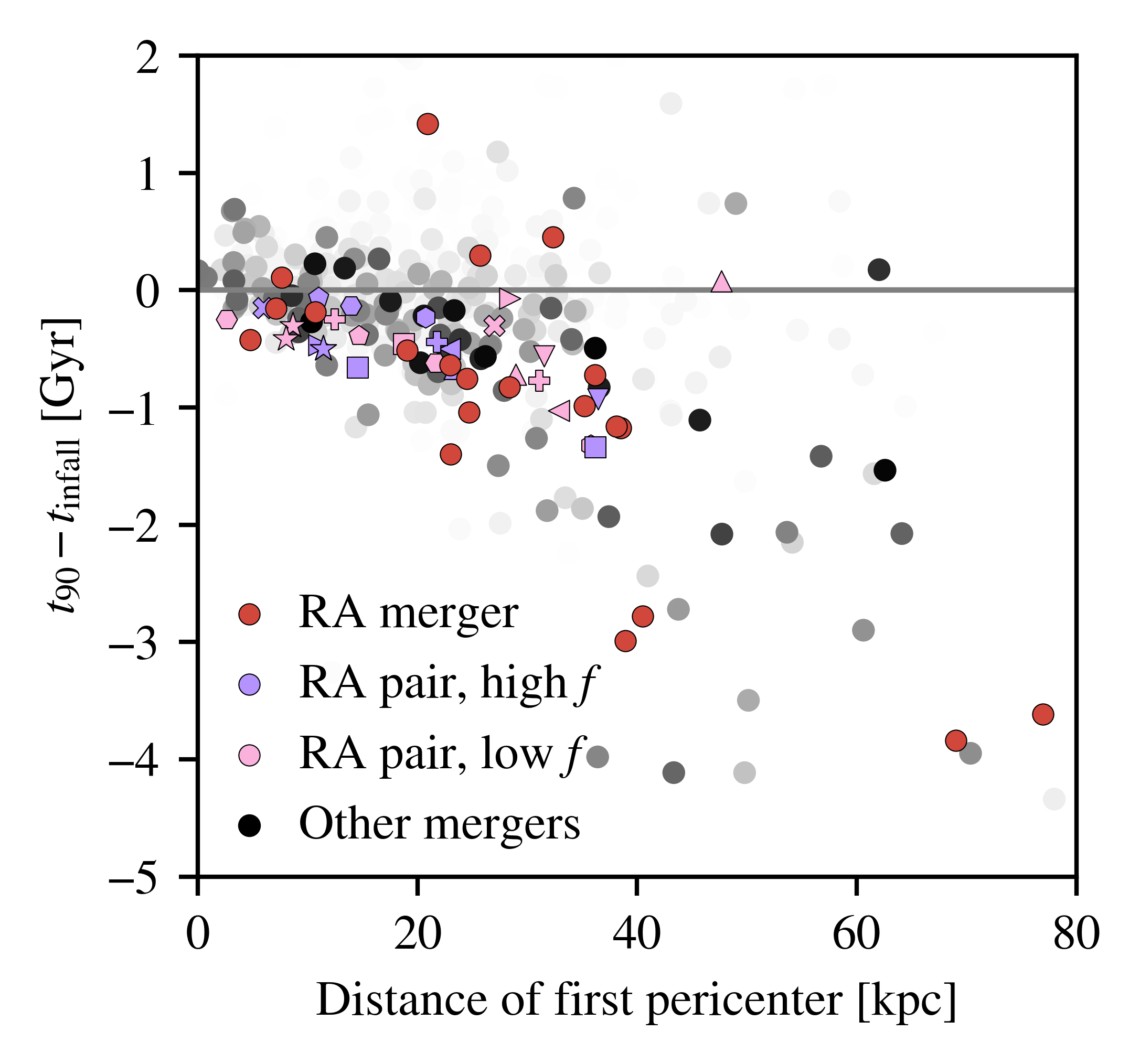}
  \caption{The quenching time $t_{90}$ relative to the infall time \tin{}, shown against the distance of first pericenter for each galaxy contributing to RA debris. All markers are consistent with \autoref{fig:3}. Galaxies that continue forming stars after crossing the virial radius are below the gray $t_{90} = \tin{}$ line. While the quenching time is typically within 1~Gyr of the infall time, it is possible for star formation to continue well after the infalling galaxy crosses the virial radius. This is especially true of galaxies on orbits with large pericentric distances: in these cases, star formation can continue for many gigayears after infall.}
  \label{fig:7}
\end{figure}

\begin{figure*}
  \centering
  \hspace{-3.5em}
  \includegraphics{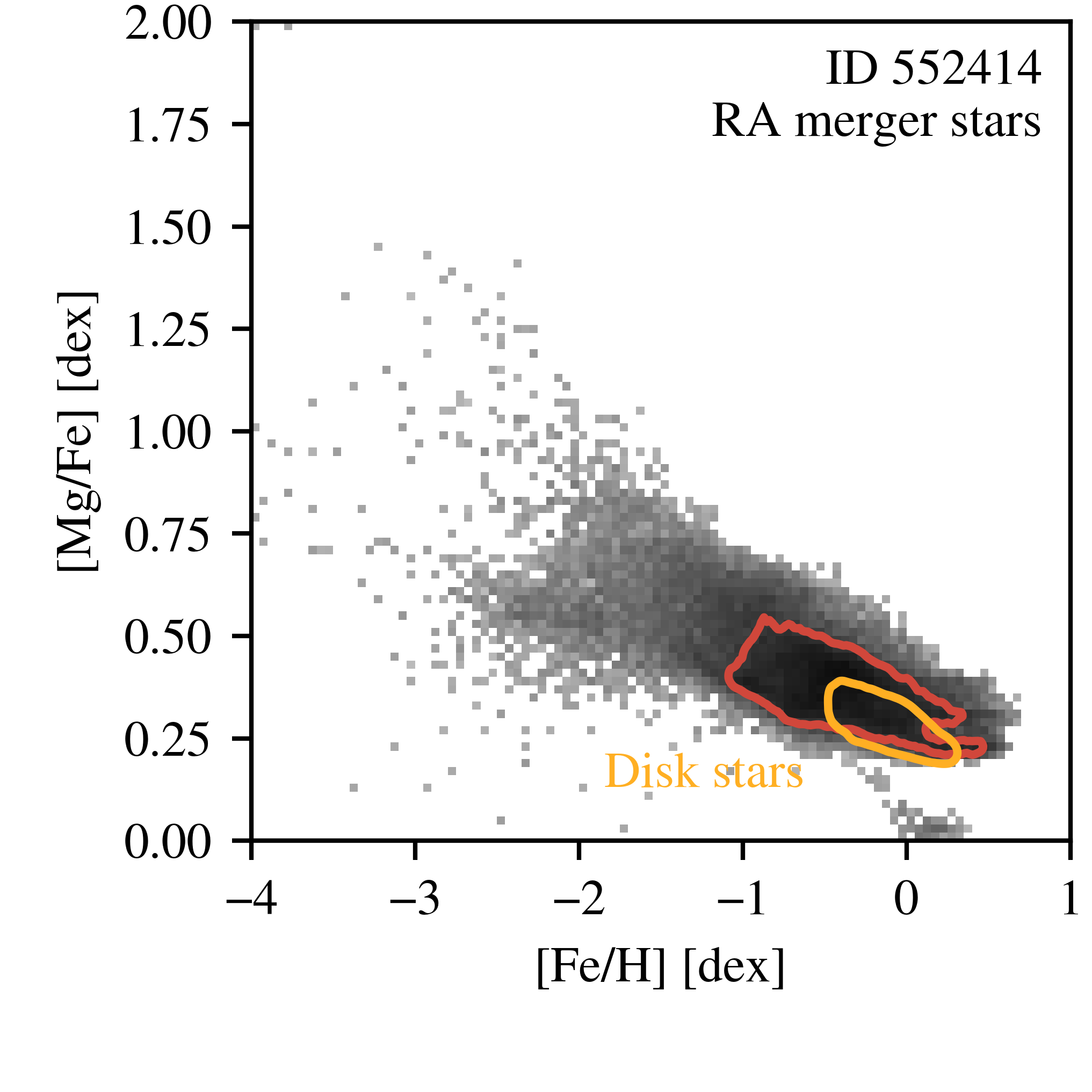}
  \includegraphics{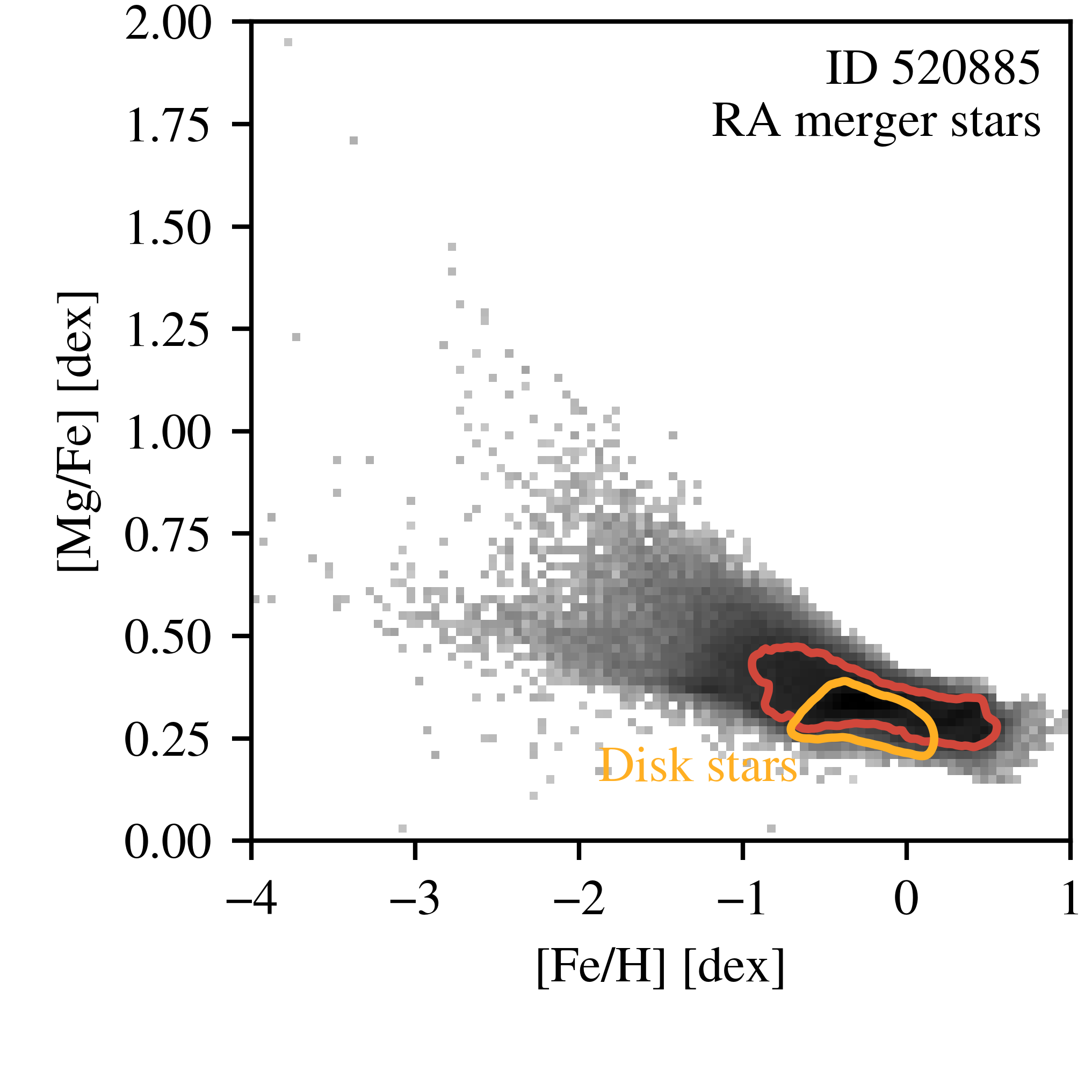}\\
  \hspace{-3.5em}
  \includegraphics{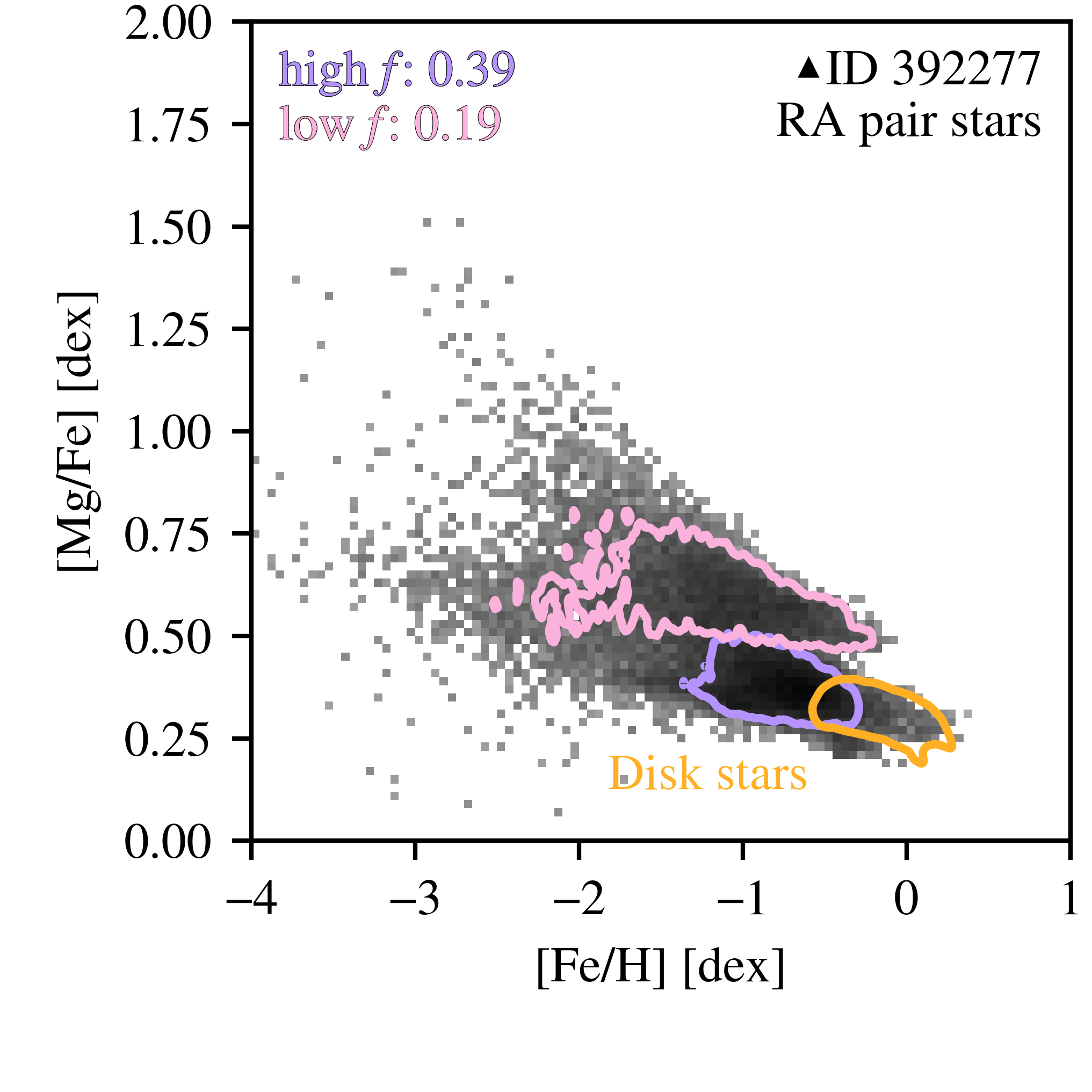}
  \includegraphics{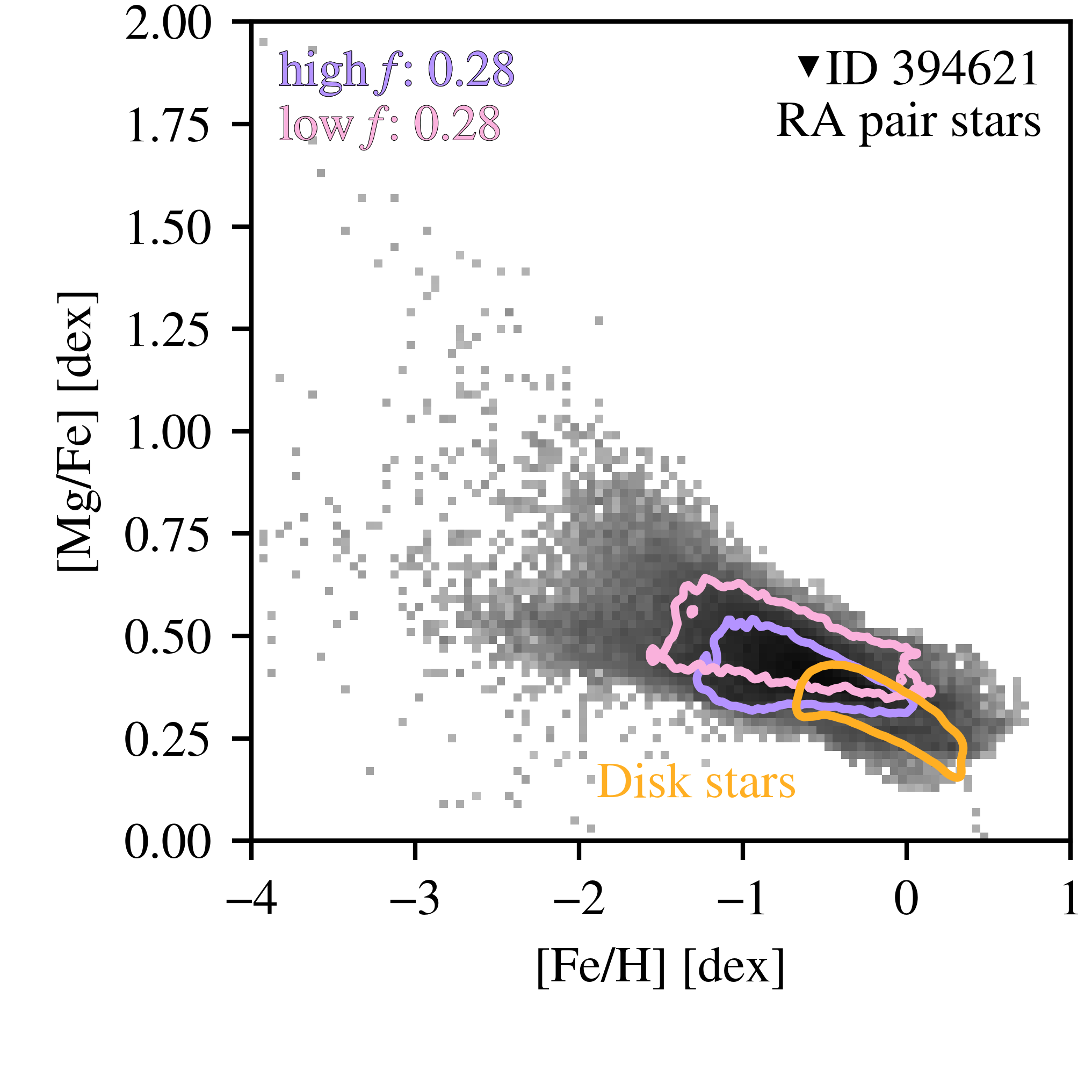}
  \caption{Tinsley--Wallerstein diagram showing [Mg/Fe] abundance relative to [Fe/H] abundance for the same RA debris shown in \autoref{fig:6}. Formatting is consistent with \autoref{fig:4} and \autoref{fig:5}. A gold contour is also included for the \ins{} disk in each MW analog to guide the eye. RA debris is generally metal-poor and $\alpha$-enhanced relative to the disk. This chemistry can in some cases distinguish the two populations of stars comprising an RA pair, e.g. in the bottom left panel, which displays a clear $\alpha$-enhancement in the high-$f$ merger relative to the low-$f$ merger. However, this is not always the case; in the right column, the galaxies comprising the pair formed their stars at the same time, and the chemical abundances prove ineffective at separating the two. Versions of this figure for each choice of RA debris are provided in \autoref{app:C}.}
  \label{fig:8}
\end{figure*}

\begin{figure*}
  \centering
  \includegraphics{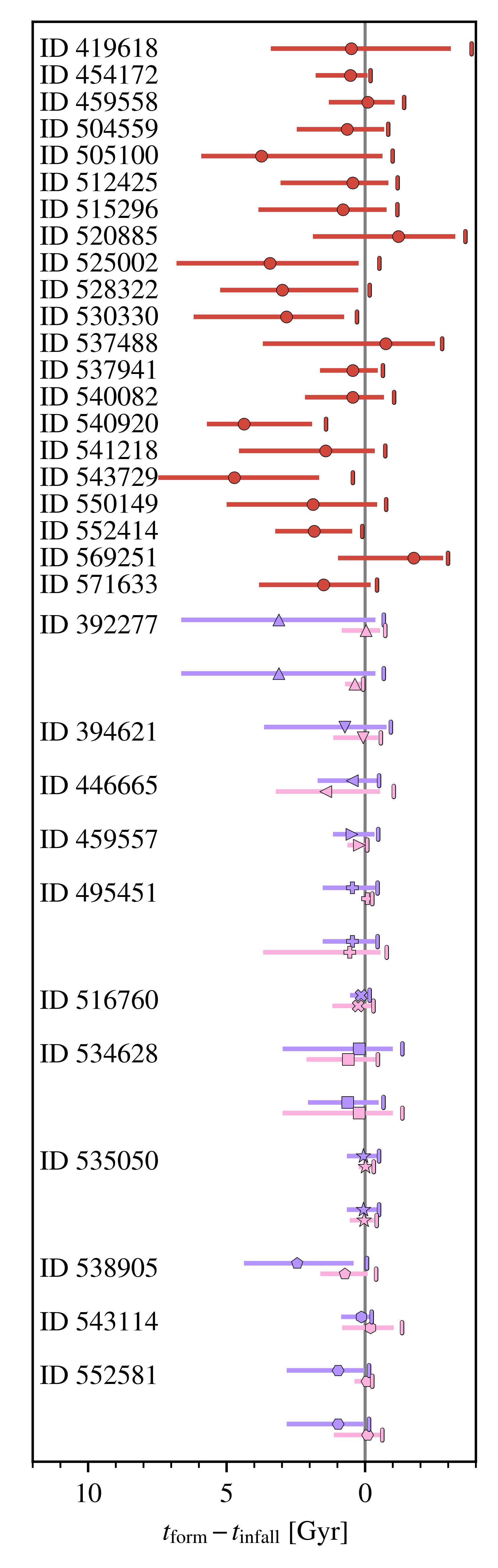}
  \includegraphics{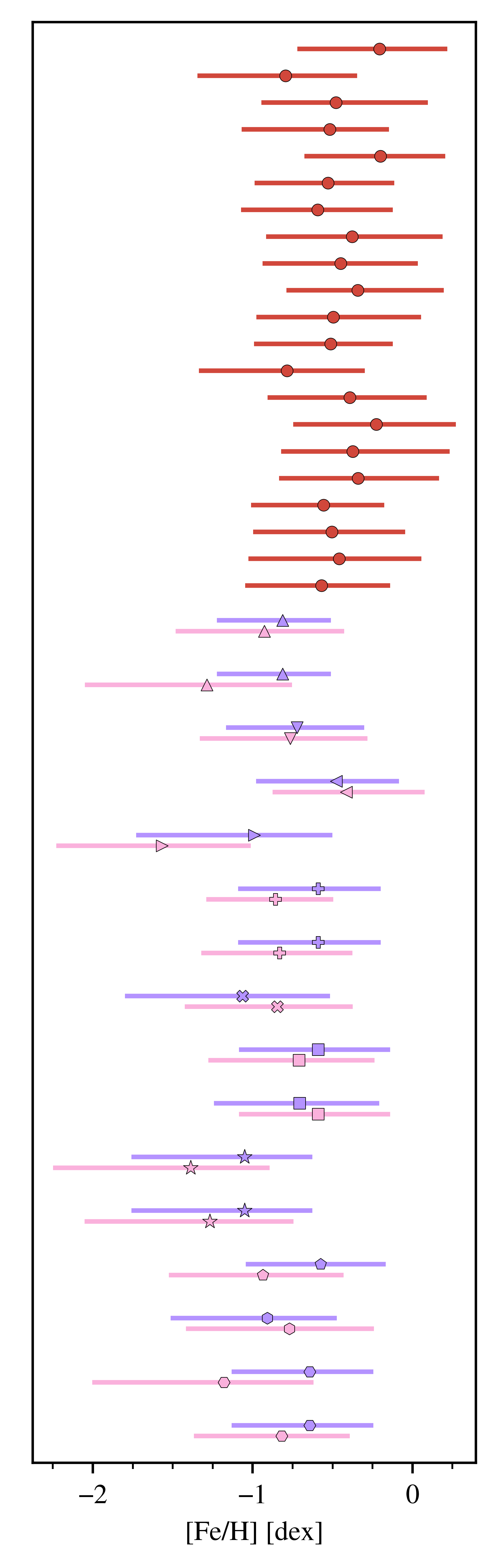}
  \includegraphics{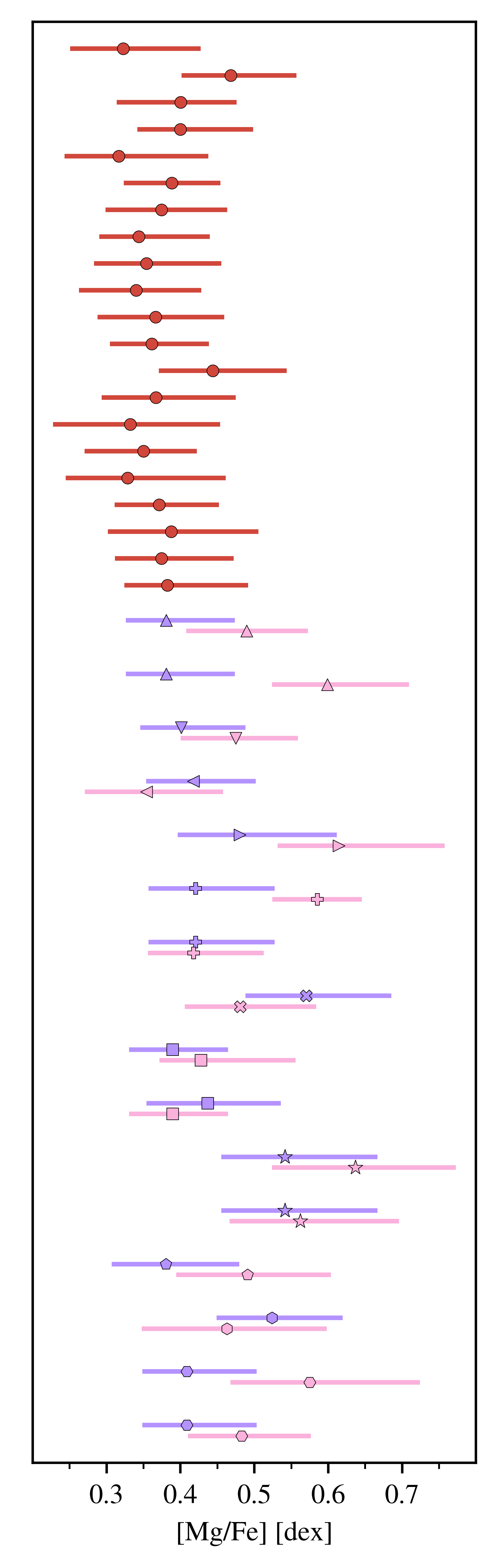}
  \caption{Comparison of quantities relevant to star formation in each of the RA debris mergers. In each panel, the highlighted region runs from the 16th to the 84th percentile for the relevant quantity, with the scatter point at the 50th percentile (with markers consistent with previous figures). The \Subfind~ID of each MW analog is also provided. (\emph{Left:})~Formation time, $t_\mathrm{form}$, of stars from each merger relative to the merger's infall time, including a colored vertical bar indicating $t_{90}$, the time at which 90\% of the galaxy's stars had formed. Generally, star formation continues for some time prior to infall, at which point it ceases, and $t_{90}\sim \tin$. As RA pairs tend to have earlier infall times and smaller masses (both in \Mdyn{} and \Mstar{}), their star formation is brief and does not extend as far prior to \tin{} as it does for RA mergers. Furthermore, RA mergers may hold on to their gas more effectively and continue forming stars many gigayears after \tin{}. (\emph{Center:})~Abundance of iron relative to hydrogen. (\emph{Right:})~Abundance of magnesium relative to iron. These right two panels show some systematic trends among the RA debris, with RA pairs more metal-poor and $\alpha$-enhanced than the single RA mergers, though these differences are less pronounced relative to the overall scatter in chemical abundances.}
  \label{fig:9}
\end{figure*}
The bottom row of \autoref{fig:5} shows the RA pair examples in orbital energy and angular momentum space. In the left panel, the contributions from each merger are more easily distinguished from each other: only 12\% of the high-$f$ merger stars fall within the pink contour, and 19\% of the low-$f$ stars fall within the purple contour. Conversely, the right panel's RA pair is not easily distinguished as two independent mergers, as 36\% of the high-$f$ merger stars fall within the pink contour, and 50\% of the low-$f$ fall within the purple contour. This overlap occurs even though the two mergers comprising the RA pair have infall times separated by more than 1~Gyr~(\autoref{tab:2}) and have totally independent orbital initial conditions; their debris still settles into the same region of this kinematic space. There are hints that such overlap occurs in observational data, making it difficult to determine whether, e.g., structures like Sequoia, Arjuna, LMS-1, and I'itoi are truly independent structures or whether they should be considered part of the GSE debris~\citep{Koppelman19,Massari19,Malhan21,Horta23}.

\Needspace*{4\baselineskip}
\section{Star Formation Histories}
\label{sec:5}
Chemodynamical space is particularly effective at separating various populations of \exs{} material, as stars from different infalling dwarf galaxies will generically have different formation environments and therefore different chemical abundances. Chemical information can also be used to infer the ages of \exs{} stars, which gives insight into the quenching time and overall SFH of the dwarf in which they originated. Because there are systematic shifts in the infall times of single mergers and pairs of mergers that contribute RA debris (see~\autoref{sec:3.3}), studying the SFH---and thence inferring the infall time of the dwarf galaxy progenitor(s)---is a promising route to distinguishing the two scenarios. 

Infalling dwarfs are thought to stop forming stars quickly, as the MW's circumgalactic medium strips them of their gas \citep{Simpson18,Akins21,Engler23}. A close look at the satellites of MW-like hosts in \TNG{} highlights the diversity of SFHs that can result from this process. In some cases---especially for the smallest, earliest-infalling satellites---star formation ceases soon after the satellite crosses the host's virial radius. In other cases, the gas is not immediately stripped, and star formation continues for some time: for example, \citet{Engler23} have shown that satellite quenching around MW analogs occurs 2.5~Gyr after infall for the median satellite in the \TNG{} model and 7.5~Gyr after infall for the 84th percentile satellite. \autoref{fig:6} illustrates the diversity across four example MWs in our sample. The horizontal axis corresponds to the formation time of \exs{} stars, $t_{\rm form}$. The black histogram is the total SFH, while the colored histograms are the contributions from the RA debris. The infall time of each merger, \tin{}, is indicated by a dashed vertical line of the same color. Each dotted vertical line corresponds to the quenching time, $t_{90}$, by which 90\% of a dwarf's stars have formed.

The left column of \autoref{fig:6} showcases galaxies in which star formation ceases quickly after infall, with $\tin{} \sim t_{90}$. Furthermore, the galaxies comprising the RA pair~(bottom left panel) are well separated in terms of the formation times of their stars, meaning the SFHs of these two galaxies are sufficient to distinguish them from each other. The right column, however, highlights cases in which this is not the case. For the single merger, in the top right panel, star formation continues for some time following the dwarf's infall, making it difficult to infer the \tin{} from the SFH. For the RA pair, in the bottom right panel, the two galaxies form stars at very similar times, with overlapping SFHs that would make distinguishing the two stellar populations difficult on the basis of their age.

Whether or not star formation continues after \tin{} is influenced by the galaxy's orbit as it is accreted onto the MW analog. In particular, \autoref{fig:7} shows that continued star formation is correlated with large pericentric distances: the horizontal axis shows the distance of the infalling galaxy's first pericenter, and the vertical axis shows the quenching time $t_{90}$ relative to the infall time $\tin$. Most galaxies---including those shown in \autoref{fig:6}, save the top right panel---cease forming stars within 1~Gyr of their infall, but this is especially true for galaxies on orbits with small pericenters. In these cases, the galaxy falls directly toward the MW analog's dense inner region, and star formation is prevented due to gas loss and tidal stripping. Galaxies with larger pericentric distances---such as ID~520885, shown in the top right panel of \autoref{fig:6}, which has a pericentric distance of 77~kpc---are less affected by stripping and are able to continue forming stars for many gigayears after their infall \citep[see][]{DiCintio21}. 

Including chemical abundances, especially for $\alpha$-process elements, reveals more detailed information about the environment in which the stars formed. \autoref{fig:8} shows a Tinsley--Wallerstein diagram for the same halos as \autoref{fig:6}, with formatting similar to previous figures: the grayscale histograms show the RA debris, with contours enclosing 68\% of the stars from each merger. For reference, a gold contour encloses \ins{} disk stars, i.e. stars bound to the MW analog at the time of their formation that have orbital circularity $j_z/j_\mathrm{circ} > 0.7$ \citep{Scannapieco09,Sotillo-Ramos22}.\footnote{Here, $j_z$ is the $z$-component of the star's specific angular momentum about the galactic center, and $j_\mathrm{circ}$ is the specific angular momentum of a circular orbit at the star's current galactocentric distance.}

While specific values of chemical abundances depend on the stellar physics model of \TNG{},\footnote{The chemical abundances are in keeping with the current state-of-the-art, but they are based on metal yields that are rather uncertain~\citep{Pillepich18}. For example, it is known that \TNG{} MW analogs are fairly $\alpha$-enriched relative to observations, see \citet{Naiman18}. \citet[\S{}2.3]{Semenov24} provide approaches to correcting the metallicity distributions of \TNG{} MWs, but in this work we instead focus on systematic shifts.} relative shifts from one merger to another (and of mergers relative to the disk) still serve as an interesting comparison point. In the bottom row of \autoref{fig:8}, the left panel shows the RA pair comprised of mergers with more distinct SFHs. Here, the two components have a clear separation in [Mg/Fe] relative to each other, with two sequences present in the histogram. In this pair, 10\% of the high-$f$ merger stars fall within the pink contour and 2\% of the low-$f$ merger stars fall within the purple contour. The bottom right panel shows the example RA pair with concurrent SFHs, and here the $\alpha$ abundance is not enough to clearly distinguish the two mergers from each other. In this pair, 54\% of the high-$f$ merger stars fall within the pink contour and 39\% of the low-$f$ merger stars fall within the purple contour. 

\autoref{fig:9} summarizes the SFHs and chemical abundances across the sample: for each galaxy contributing to RA debris, a marker and horizontal line show the median and spread of the relevant quantity. The left panel shows the duration of star formation with respect to the infall time of the mergers, with an additional marker for the quenching time, $t_{90}$, at which 90\% of the merger's stars are formed. The duration of star formation is one suggestive distinction between RA mergers~(in red) and RA pairs~(in purple and pink), as the former have SFHs that extend well before their \tin{}, while the latter form their stars quickly, as they are generally accreted earlier in cosmic history~(see~\autoref{fig:3}). The panel also reveals that $t_{90}$ generally serves as a good proxy for \tin{}; for most galaxies in this sample, the $t_{90}$ marker is close to the line through the origin that denotes the infall time. Notably, the exceptions to this trend are often single RA mergers, rather than the mergers of an RA pair; in terms of inferring \tin{} from $t_{90}$, this further biases RA mergers toward later-inferred infall times.

The clear differences in SFH are less concretely reflected in chemical abundances in the \TNG{} model, though two systematic shifts are still present. First, in [Fe/H], RA pairs are biased in the metal-poor direction relative to single RA mergers, indicative of the early epochs in which their stars formed. Second, in [Mg/Fe], the RA pairs are biased toward $\alpha$-enhancement, especially for the smaller-$f$ component of the pair (shown in pink). While these trends may not be sufficient to distinguish individual mergers from pairs (or galaxies comprising pairs from each other), the existence of the systematic biases between populations serves as a potential indicator for the merger histories of MW analogs. This is especially true if stellar ages can be robustly determined, as these trends are more apparent in SFH than in stellar chemistry.

\Needspace*{4\baselineskip}
\section{Conclusion}
\label{sec:6}
This work presents a sample of 98 MW analogs in the \TNGfifty{} simulation of the \Illustris\TNG{} suite. Each analog has its full accretion history identified with each \exs{} star assigned to the merger that contributed it (\autoref{sec:2.3}). This is the largest sample of MW analogs for which this decomposition has been performed, and it allows for unprecedented statistics in studying the assembly of the MW and the buildup of its stellar halo (\autoref{sec:3.1}).

The MW analogs are chosen to match some properties of the Galaxy, namely its stellar mass and relative isolation (\autoref{sec:2.2}), but certain aspects of the MW itself are rare across the sample. In particular, no analogs have an LMC-like satellite with mass $\Mdyn > 10^{11}~\Msun$, indicating that the presence of the LMC distinguishes our Galaxy as an unlikely system.\footnote{Recall that we make no requirement that the analogs have companion galaxies resembling the LMC, M31, or other Local Group objects. Additionally, while the MW analogs are more than 1~Mpc away from any cluster, their large-scale environments do not in general match that of our Galaxy. See \autoref{sec:2.2} for more details.} Further, while 33\% of MW analogs have stars with properties qualitatively similar to the GSE debris seen in our Galaxy---namely, highly radially biased velocities comprising a majority of the inner \exs{} stars~(\autoref{sec:3.2})---only 4\% of MWs have debris as anisotropic as the GSE debris. This is in line with other studies that find the MW to be rare: see, e.g., \citet{Evans20,Buch24}.

With the large sample of MW analogs, we can uncover the origins of GSE-like debris in a $\Lambda$CDM context. In particular, we consider scenarios in which the GSE debris is comprised of multiple independent mergers, rather than being accreted all at once. Approximately a third of the GSE-like debris seen in \TNG{} is built up from two mergers. Statistics alone, therefore, do not preclude the possibility that the GSE debris in our own Galaxy is comprised of such a pair of mergers.

Given the possibility of a two-merger GSE, the primary concern is in distinguishing it from a single-merger origin; however, it proves difficult for observable parameters to differentiate the two scenarios. Kinematics alone provide little conclusive evidence, as shown in \autoref{sec:4}. Single mergers can take on relatively smooth, featureless distributions in kinematic spaces, but they can also exhibit complex structures with multiple overdensities in kinematic parameters \citep[see][]{Jean-Baptiste17,Belokurov23}. Merger pairs, too, can also lead to well-separated stellar debris that is fairly easily decomposed into the individual dwarf galaxy components, but may also yield kinematic distributions in which the galaxies comprising the merger are indistinguishable, occupying the same regions of these kinematic spaces.

Though we find it difficult to distinguish stars' origins in 2D projections of kinematic spaces, machine learning techniques may be able to take advantage of the high dimensionality of the available parameter space to effectively classify stars according to their galactic progenitors. Such an algorithm would require a large amount of training data. This study is already a step toward a large statistical sample of GSE-like events, and as simulations continue to progress, sample sizes will continue to grow. 

Furthermore, though kinematic spaces such as $v_r - v_\phi$ and $E-L_z$ are ineffective at distinguishing merger events from one another, other parameters such as orbital frequencies may hold more discerning power \citep{Gomez10}. This requires detailed modeling of the MW's gravitational potential, and efforts in this regard are well underway \citep[e.g.][]{Garavito-Camargo21,Reino21,Palau23,Ibata24}. Further, the GSE itself influences the MW's potential, which has observational consequences~\citep[e.g.][]{Han22,Han23,Han23a}. Including these efforts into the modeling of GSE debris will further aid in understanding its origin and composition. 

Crucially, SFHs and chemical abundances \emph{do} uncover systematic shifts between single-merger GSE debris and two-merger GSE debris, shown in \autoref{sec:5}. Single mergers are typically accreted later than the pairs, and correspondingly they have larger masses (both in terms of their stellar mass and their total mass), a longer duration to their star formation, and a slight metal enrichment relative to the pairs. These general trends do not, however, necessarily imply that these cases are observationally distinguishable, especially given that the largest shifts are seen in the distributions of stellar ages, which have observational uncertainties on the order of gigayears \citep[e.g.][]{Chaplin14,Gallart19,Grunblatt21,Johnson23}. 

Determining the details of the GSE's SFH is, therefore, a productive step in understanding its true nature. The literature has progressed in this regard, with many studies \citep[e.g.][]{Helmi18,Gallart19,Bonaca20,Feuillet21,Grunblatt21,Hasselquist21,Montalban21,Johnson23,Ernandes24,Horta24} determining the ages of GSE stars and increasingly detailed accounts of GSE star formation. The general consensus of this body of work is that the peak of GSE star formation was $\sim 10$~Gyr ago, with a tail that continues for $\gtrsim 2$~Gyr as the dwarf fell into the MW and slowly stopped forming stars. The extended duration of star formation is consistent with a gradual infall with large pericenter---see~\autoref{fig:7} and the simulations of \citet{Naidu21}---or multiple smaller mergers---see~\citet{Donlon22,Orkney23,Rey23}. There is still more work to be done in this direction: as a greater number of stars have their ages robustly determined, the time resolution of GSE star formation will be improved, giving deeper insight into the extragalactic source of this debris. 

\section*{Acknowledgments}
The authors would like to acknowledge Andrea Caputo, Andreia Carrillo, Akaxia Cruz, James Johnson, Kassidy Kollmann, Matthew Orkney, Jonah Rose, and Risa Wechsler for helpful conversations. D.F. and M.L. are supported by the Department of Energy~(DOE) under award number DE-SC0007968. M.L. is also supported by the Simons Investigator in Physics Award. This research was supported in part by grant NSF PHY-2309135 to the Kavli Institute for Theoretical Physics~(KITP). L.N. is supported by the Sloan Fellowship, the NSF CAREER award 2337864, and NSF award 2307788. L.H. acknowledges support by the Simons Collaboration on ``Learning the Universe.'' The computations in this paper were run on the FASRC cluster supported by the FAS Division of Science Research Computing Group at Harvard University. The \Illustris\TNG{} simulations were undertaken with compute time awarded by the Gauss Centre for Supercomputing~(GCS) under GCS Large-Scale Projects GCS-ILLU and GCS-DWAR on the GCS share of the supercomputer Hazel Hen at the High Performance Computing Center Stuttgart~(HLRS), as well as on the machines of the Max Planck Computing and Data Facility~(MPCDF) in Garching, Germany.

\software{
  astropy \citep{AstropyCollaboration22}, 
  bottleneck \citep{Goodman16}, 
  h5py \citep{Collette13},
  Jupyter \citep{Kluyver16},
  matplotlib \citep{Hunter07},
  numba \citep{Lam15},
  numpy \citep{Harris20},
  SciPy \citep{Virtanen20}, 
  TNG access scripts \citep{Nelson21}
}

\FloatBarrier

\appendix
\section{The Milky Way Analogs}
\label{app:A}

The snapshot-99 \Subfind~IDs of the 98 Milky Way~(MW) analogs studied in this work are given below. The criteria by which these are selected are provided in \autoref{sec:2.2}.

\begin{quote}
  \texttt{[ 5, 7, 9, 208812, 372755, 392277, 394621, 419618, 444134, 446665, 449658, 454172, 455291, 459557, 459558, 476266, 477328, 478216, 479938, 480802, 490079, 490814, 493433, 495451, 497557, 499704, 502371, 503987, 504559, 505100, 509091, 510585, 511303, 512425, 513105, 515296, 516101, 516760, 517271, 518682, 520885, 521429, 521803, 522530, 522983, 523889, 524506, 525002, 526478, 528322, 528836, 529365, 530330, 530852, 531910, 532301, 532760, 534628, 535050, 535410, 535774, 537236, 537488, 537941, 538905, 539333, 540082, 540452, 540920, 541218, 541497, 541847, 542252, 543114, 543376, 543729, 544001, 544408, 545437, 546114, 546474, 546870, 547293, 547844, 549516, 550149, 552414, 552581, 554798, 555601, 557721, 559386, 560751, 569251, 571454, 571633, 572328, 613192]}
\end{quote}
\FloatBarrier

\section{Technical Details for Determining Accretion Histories}
\label{app:B}
\renewcommand{\thefigure}{B\arabic{figure}}
\setcounter{figure}{0}

The following describes the modifications made to \Sublink{} to obtain the merger tree for the MW analogs~(\autoref{sec:B1}) and the details of how stars are associated with mergers~(\autoref{sec:B2}). Together, these provide implementation details for the methods described in~\autoref{sec:2}.

\subsection{MW Merger Trees}
\label{sec:B1}
\begin{figure*}[t!]
  \centering
  \includegraphics{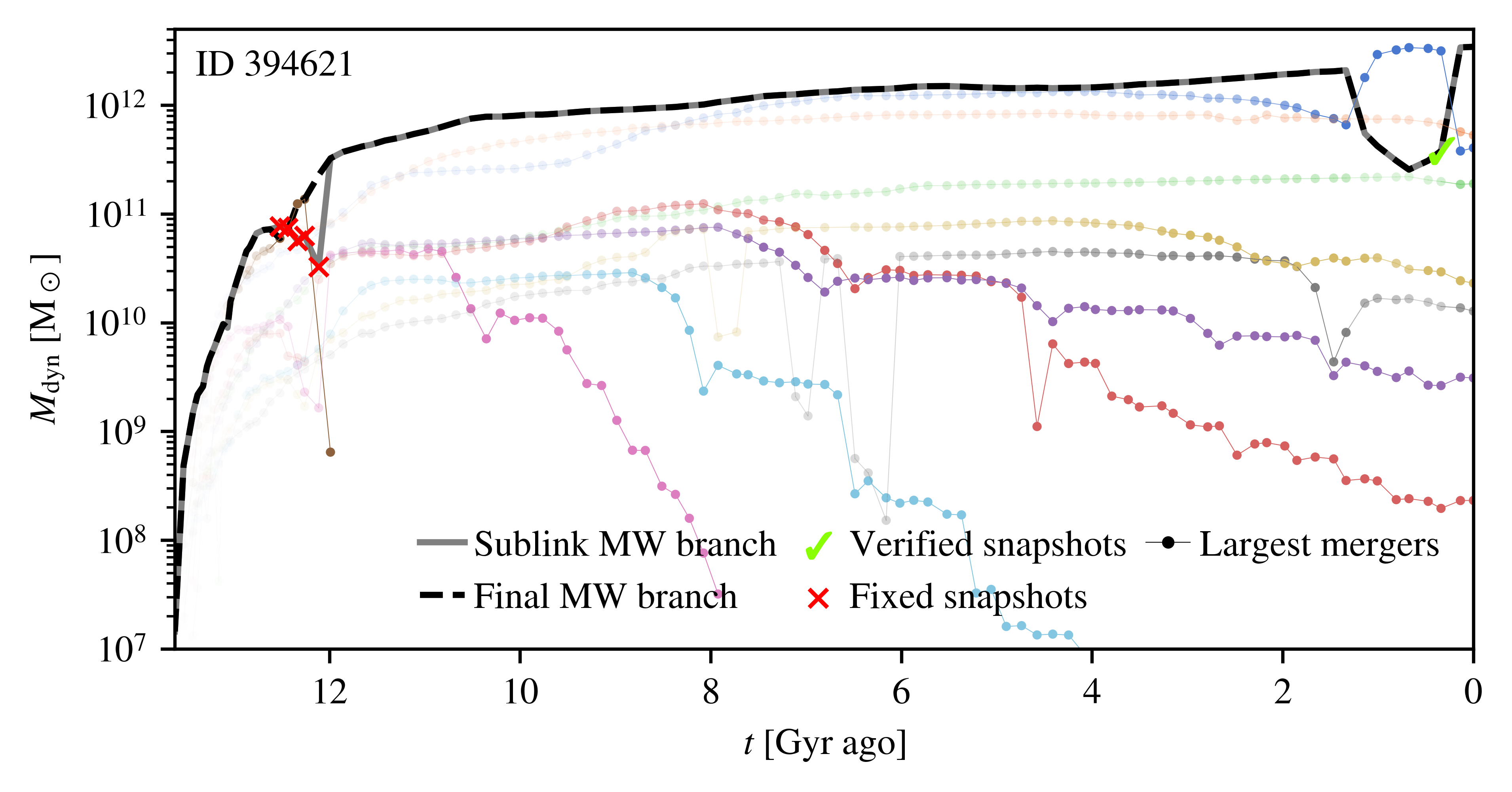}
  \caption{An example merger tree for one MW analog with ID~394621. The horizontal axis is the time corresponding to each snapshot, and the vertical axis is the mass of the halos in the merger tree at each snapshot. The MW analog's branch from \Sublink{} is shown in solid gray, and the patched branch used in this work is in dashed black. The largest mergers are shown with colored markers, where the opacity corresponds to the distance from the host halo: the markers are opaque when the merger is within the MW analog's \Rdyn{}. The snapshots on the \Sublink{} branch marked with a green check mark are those that are verified as correct (see text), while the snapshots marked with a red cross are those where the halo ID used in this work is changed from the ID provided by \Sublink{}.}
  \label{fig:B1}
\end{figure*}

The snapshot-by-snapshot history of the MW analogs can be largely determined via the \Sublink{} algorithm~\citep{Rodriguez-Gomez15} and corresponding ``merger tree'' catalogs provided by~\cite{Nelson19}. At a given snapshot, \Sublink{} identifies candidate ``descendant'' \Subfind{} halos in the subsequent snapshot that share particles with this ``progenitor'' halo. The unique choice of descendant---which is, ideally, either the same physical structure as the progenitor or a larger halo into which the progenitor merges---is chosen from this list of candidates, with priority given to those halos that contain the most-bound particles from the progenitor. 

From this process, a full tree is built up, with ``branches'' consisting of a halo, its descendants in subsequent snapshots, and its progenitors in previous snapshots. An example of such a tree is shown in \autoref{fig:B1}, in which the host halo (i.e., the MW analog, henceforth simply called ``the MW'' for ease of discussion) has a branch determined by \Sublink{} that is shown in gray, and the largest mergers are shown in various colors, with the opacity indicating distance from the MW; the markers are fully opaque when the merger is at or within \Rdyn{}. This MW experienced a merger $\sim 1.5$~Gyr ago with a halo (the darker blue markers) that has a mass comparable to that of the MW itself. Such a violent merger is difficult to resolve, both in terms of \Subfind{} isolating the individual halos from each other and further in terms of \Sublink{} connecting the halos from each snapshot robustly into a merger tree. In this case, the halos of the MW branch dip in mass by over 1~dex, while the merging object increases in mass by a similar amount. At an earlier point in the MW's evolution, approximately 12~Gyr ago, another similarly large merger occurred. In this case, the MW halos identified by \Subfind{} again experience a jump in \Mdyn{} of approximately 1~dex.

Jumps in halo mass that occur this rapidly may be indicative of cases where many particles previously bound to the halos of the MW branch instead get attributed to the halos of the merger's branch. To mitigate these cases, we modify the MW branch: if the total bound mass \Mdyn{} has a significant~($> 0.2$~dex) increase between two snapshots, this is taken as an indication that such a misidentification has taken place. We assume that the higher-mass halo on the MW branch, in the snapshot closer to $z = 0$, is correctly identified and then double-check that the lower-mass halo in the prior snapshot is its true progenitor. This is done by taking the central black hole in the descendant halo and finding it in the preceding snapshot. The halo that contains the black hole in this snapshot is then labeled the main progenitor and used in place of the halo identified by \Sublink{}. This process then recurses, starting at the most recent spurious snapshots and working backward through time until either there are no such jumps in the mass or there is no longer a black hole present. 

The result of this process is shown in \autoref{fig:B1}: at the right edge of the figure, between the third-to-last and second-to-last snapshots, the MW halo experiences a large jump in mass. A black hole is found in the second-to-last snapshot, and we confirm that it resides in the MW halo from the preceding snapshot. This is indicated by the green check mark and the agreement between the \Sublink{} tree (in gray) and the amended MW branch (in black). This case highlights the difficulties in merger tree algorithms: it is unlikely that the MW halo should truly be considered to have lost this much mass, but faithfully tracking the physical structures with a halo-finding algorithm is complicated, especially in large-mass-ratio mergers, and linking these halos together into a merger tree branch can be challenging. After this jump, the algorithm then recurses and finds no further jumps until very early on in cosmic history, at $t\sim 12$~Gyr ago. At this jump, a black hole exists and is used to re-select the MW halo. In this case, our algorithm rejects the low-mass halo found by \Sublink{} (indicated with the red cross). Instead, we find the black hole in a different halo, one that \Sublink{} considers to be a merger rather than the MW itself. We use this halo as the MW rather than the halo chosen by \Sublink{}. This process repeats for a total of five snapshots, where in each case the \Sublink{} halo does not contain the MW's black hole, so it is not used as part of the MW branch. These alterations result in a smoother mass evolution of the MW, without the sharp jump seen in the \Sublink{} definition branch. 

\subsection{Associating Stars to Mergers}
\label{sec:B2}
With the MW's branch properly identified within the merger tree as described above, the other branches of the tree contain only the accreted halos. Infalling mergers are identified first by considering the MW's merger tree. They are given a unique history by following the branches from early snapshots (starting at every ``main leaf progenitor'' in the tree) toward $z = 0$ until either the MW host branch is reached (i.e., the infalling halo is completely merged with the MW) or the \Sublink{} descendant halo otherwise has a different main progenitor (i.e., the infalling halo has merged with some other object). 

This process will not in general fully characterize the merger history: halos that survive to $z = 0$ will never have the MW as a descendant halo and therefore may not be in the MW's tree. Therefore, we extend the \Sublink{} tree using the data collected by tracking the individual stars (\autoref{sec:2.3}). We first identify stars that are not born in the MW analog but that are never bound to any merger halo identified by the above procedure. We take \Sublink{} trees for the halos that host these stars and isolate the branches corresponding to them following the above procedure, ensuring that the branch is tracked until merging with a larger halo, i.e. until the descendant halo has a different main progenitor. 

With the MW analog's \Sublink{} tree and the additional merger information added from the stars themselves, the full merger history of the MW analog is known. As mentioned in the main text, stars are assigned to the merger to which they were bound for the greatest number of snapshots. In ambiguous cases, where a star spent the same number of snapshots in multiple mergers, preference is given to the merger with the largest peak mass. There is ambiguity for $1.25_{-1.08}^{+5.76}\%$ of \exs{} stars in a given MW, and typically ($83_{-29}^{+15}\%$ of the time) these stars are found in the branches for only one snapshot each; these edge cases of ambiguous merger assignment are generally not stars that spend long periods of time bound to a merging halo. 

Recall that we allow halos previously assigned to merger branches of the tree to be reassigned to the MW's branch, which can happen when the MW halo momentarily has far fewer particles than in the subsequent snapshot. This happens especially in cases of mergers with large mass ratios and small pericentric distances, when particle assignment to halos and halo assignment to merger branches is difficult to establish. When this occurs, a single \Sublink{} branch may be considered two mergers: the original branch is thought to be fully merged at pericenter, as its descendant halo is the MW, and if the merging halo is re-identified after its pericenter, this would be considered a second merger. A similar effect occurs if \Sublink{} fails to connect halos between snapshots that are physically the same object. We checked that none of the RA pairs identified in this work are affected by this issue; they are all genuinely independent mergers. 

\FloatBarrier
\section{Additional Figures}
\label{app:C}
\renewcommand{\thefigure}{C\arabic{figure}}
\setcounter{figure}{0}

The main body of this work highlights a few examples of RA debris in terms of kinematics (see \autoref{fig:4}, which shows the distribution of $v_r$ versus $v_\phi$ for a selection of RA debris, and \autoref{fig:5}, which shows the distribution of $E$ versus $L_z$ for the same mergers) and in terms of star formation history (see \autoref{fig:6}) and chemical abundances (see \autoref{fig:8}). For completeness, the rest of these distributions are provided online.\footnote{\href{https://github.com/folsomde/Stellar_Halos_with_GSEs}{\texttt{https://github.com/folsomde/Stellar\_Halos\_with\_GSEs}}, or on Zenodo under an open-source Creative Commons Attribution license: \dataset[\texttt{doi:10.5281/zenodo.14829722}]{https://doi.org/10.5281/zenodo.14829722}.}

% \input{AppC}

% \bibliographystyle{aasjournal}
% \bibliography{main}
\bibliographystyle{aasjournal}
\bibliography{main}

\end{document}